\documentclass[%
superscriptaddress,
reprint,
 amsmath,amssymb,
prb
]{revtex4-1}
\usepackage{color}
\usepackage{graphicx}
\usepackage{dcolumn}
\usepackage{bm}

\begin{document}

\title{In situ STM observation of nonmagnetic  impurity effect  in MBE-grown CeCoIn$_5$ films}

\author{Masahiro Haze}
\email{haze.masahiro.2x@kyoto-u.ac.jp}
 \affiliation{Department of Physics, Kyoto University, Sakyo-ku, Kyoto 606-8502, Japan}
\author{Yohei Torii}
 \affiliation{Department of Physics, Kyoto University, Sakyo-ku, Kyoto 606-8502, Japan}
\author{Robert Peters}
 \affiliation{Department of Physics, Kyoto University, Sakyo-ku, Kyoto 606-8502, Japan}
\author{Shigeru Kasahara}
 \affiliation{Department of Physics, Kyoto University, Sakyo-ku, Kyoto 606-8502, Japan}
\author{Yuichi Kasahara}
 \affiliation{Department of Physics, Kyoto University, Sakyo-ku, Kyoto 606-8502, Japan}
\author{Takasada Shibauchi}
 \affiliation{Department of Advanced Materials Science, University of Tokyo, Kashiwa, Chiba 277-8561, Japan}
\author{Takahito Terashima}
 \affiliation{Department of Physics, Kyoto University, Sakyo-ku, Kyoto 606-8502, Japan}
\author{Yuji Matsuda}
 \affiliation{Department of Physics, Kyoto University, Sakyo-ku, Kyoto 606-8502, Japan}

\newcommand{\mh}[1]{{\color{red} #1}}
\newcommand{\cm}[1]{}

\date{\today}

\begin{abstract}

Local electronic effects in the vicinity of an impurity provide pivotal insight into the origin of unconventional superconductivity, especially when the materials are located on the edge of magnetic instability. In high-temperature cuprate superconductors, a strong suppression of superconductivity and appearance of low-energy bound states are clearly observed near nonmagnetic impurities. However, whether these features are common to other strongly correlated superconductors has not been established experimentally.   Here, we report the {$in$} {$situ$} scanning tunneling microscopy observation of electronic structure around a nonmagnetic Zn impurity in heavy-fermion CeCo(In$_{1-x}$Zn$_x$)$_5$  films,  which are epitaxially grown by the state-of-the-art molecular beam epitaxy technique in ultrahigh vacuum.  The films have very wide atomically flat terraces and Zn atoms residing on two different In sites are clearly resolved.      Remarkably, no discernible change is observed for the superconducting gap at and around the Zn atoms.   Moreover,  the local density of states around Zn atoms shows little change inside the hybridization gap between $f$- and conduction electrons, which is consistent with calculations for a periodic Anderson model without local magnetic order. These results indicate that no nonsuperconducting region is induced around a Zn impurity and do not support  the scenario of antiferromagnetic droplet formation suggested by indirect measurements in Cd-doped CeCoIn$_5$. These results also highlight a significant difference of the impurity effect between cuprates and CeCoIn$_5$, in both of which $d$-wave superconductivity arises from the non-Fermi liquid normal state near antiferromagnetic instabilities.

\end{abstract}

\pacs{Valid PACS appear here}
\maketitle


\section{Introduction}

Unconventional superconductivity often emerges from the competition between magnetically ordered and paramagnetic phases in a wide variety of strongly correlated materials, including cuprate, iron-pnictide, and heavy-fermion compounds.   In these materials, superconductivity is frequently observed close to an antiferromagnetic (AFM)  quantum critical point (QCP) where the ordered and disordered states are nearly degenerate \cite{Mathur98,Shibauchi13,Keimer15}.    It is widely believed that the proliferation of quantum fluctuations in the vicinity of the AFM QCP gives rise to the  non-Fermi liquid behavior, which is visible in many physical quantities in the normal state, and  induces the magnetically mediated pairing interaction.    To gain deep insight into the pairing mechanism and competing electronic correlations in these unconventional superconductors, more and more efforts have been focused on understanding the changes of the electronic structure produced by an impurity \cite{Balatsky06,Alloul09}.    In Bardeen-Cooper-Schrieffer superconductors with $s$-wave symmetry, a nonmagnetic impurity acts as a weak scatterer because the time reversal symmetry is preserved unlike a magnetic impurity.  On the other hand, in unconventional superconductors, both magnetic and nonmagnetic impurities can act as  pair breakers and strong scatterers.  The effect of a nonmagnetic impurity is particularly intriguing in unconventional superconductors near the magnetic instability, because the different types of orders compete and coexist in a delicate balance. The induced disorder may have the power to tip the balance in favor of one of the orders.    In fact,  remarkable phenomena caused by impurities have been reported in cuprates, where the superconductivity is strongly suppressed by a nonmagnetic Zn impurity.  The bound states around a Zn atom are most spectacularly seen in scanning tunneling microscopy (STM) \cite{Pan00}.   Furthermore, nuclear magnetic resonance (NMR) studies reveal induced local magnetic moments around Zn atoms \cite{Mahajan94,Julien00}.  On the other hand, a magnetic Ni impurity does not strongly suppress the superconductivity, although it forms impurity bound states \cite{Hudson01}.

\begin{figure}[t]
	\begin{center}
		\includegraphics[width=1\linewidth]{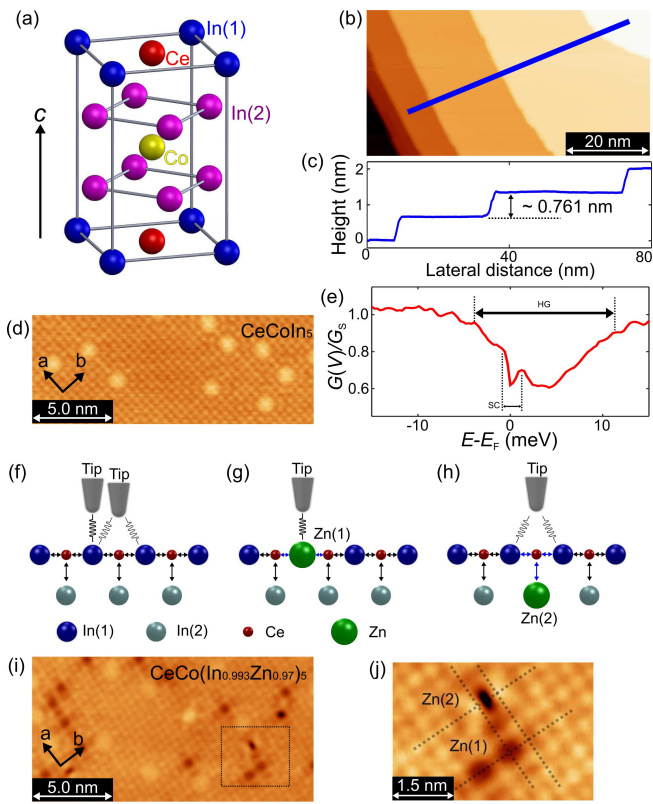}
	\end{center}
	\caption{ (a) Crystal structure of CeCoIn$_5$.  (b) STM topograph of non-doped CeCoIn$_5$. (c) Cross-sectional profile taken along the blue line in (a). (d) Topographic image of non-doped CeCoIn$_5$ with atomic resolution. Tunnel parameters: sample bias voltage $V_{\rm{S}}$ = 20 mV and tunneling current $I_{\rm{T}}$ = 200 pA.  (e) Tunneling conductance normalized by $I/V$ obtained on non-doped CeCoIn$_5$ at 0.5\,K. Tunnel parameters: $V_{\rm{S}}$ = 30\,mV, $I_{\rm{T}}$ = 100\,pA, modulation amplitude $V_{\rm{MOD}}$ = 300\,$\mu$V, and modulation frequency $f_{\rm{MOD}}$ = 1\,kHz. $\Delta_{\rm{HG}}$ and $\Delta_{\rm{SC}}$ represent hybridization and superconducting gaps, respectively. (f)-(h) Schematics of tunneling current taken above In site far away from impurities (f), Zn(1) impurity (g), and Zn(2) impurity (h), respectively. (i) Topographic image of CeCo(In$_{0.993}$Zn$_{0.007}$)$_5$ with atomic resolution. Tunnel parameters: $V_{\rm{S}}$ = 50\,mV and $I_{\rm{T}}$ = 500\,pA.  Dark spots represent Zn atoms. (j)  Topographic image of the boxed area in (i). Tunnel parameters: $V_{\rm{S}}$ = 50\,mV and $I_{\rm{T}}$ = 100\,pA. Dashed lines represent lineup of In(1) atoms on Ce plane.  There are two Zn-sites; Zn(1) residing on In(1) site and Zn(2) atom on In(2) site in subsurface. Zn(2) site is located in-between two In(1) atoms. }
\end{figure}

Among the heavy fermion systems, CeCoIn$_5$, whose crystal structure is illustrated in Fig.\,1(a),   is one of the most fascinating materials \cite{Thompson12}, as it shares many of the characteristics of cuprates, including quasi-two-dimensionality, proximity to antiferromagnetism,  and $d_{x^2-y^2}$-wave pairing superconductivity  \cite{Izawa01,Zhou13,Allan13} arising from a non-Fermi-liquid normal state.  The impurity effects on CeCoIn$_5$ have aroused a great interest because of their highly unusual features, especially when a hole is doped by substituting Cd, Hg, and Zn for In \cite{Pham06,Yokoyama15,Booth09}.   A slight doping of Cd, Hg, and Zn into CeCoIn$_5$ can change the ground state sufficiently to induce antiferromagnetism; the electronic state can be tuned between superconductivity and AFM order with a coexistence region for intermediate dopings. This behavior contrasts with that achieved by electron doping with Sn substitution for In in CeCo(In$_{1-x}$Sn$_x$)$_5$ \cite{Gofryk12,Bauer06,Romas10} and substituting with La on the Ce site in (Ce$_{1-x}$La$_x$)CoIn$_5$, in which superconductivity is suppressed but no AFM order is  observed \cite{Paglione07,Petrovic02,Bauer11}.    These results raise the crucial question about the impact of a nonmagnetic impurity in the vicinity of AFM QCP, which is of vital importance  for understanding the interplay between superconductivity and antiferromagnetism.  There are two scenarios.  First, the Fermi energy is shifted by doping holes, so that either superconductivity or AFM order emerges, depending on the competition between Ruderman-Kittel-Kasuya-Yosida interaction and Kondo effect, in much the same way that applying pressure changes the ground state.  The second scenario is that, similar to cuprates, nonmagnetic impurities  induce a local magnetic moment \cite{Mahajan94,Julien00,Neto00,Millis01,Figgins11}.  When  AFM droplets overlap with each other as the impurity concentration is increased, the system can undergo a transition into a long-range AFM state.   NMR measurements of Cd-doped CeCoIn$_5$ support the latter scenario; Cd impurity nucleates local AFM droplets \cite{Urbano07,Seo14,Sakai15}.  However, it is not clear whether the formation of the AFM droplets is a universal property caused by the substitution of a hole dopant at the In sites in CeCoIn$_5$. To provide a decisive answer to this issue,  information about the local electronic structure around a single nonmagnetic impurity is crucially important.

STM provides detailed information about the local electronic structure.   However, until now, STM measurements have been performed on only a few heavy fermion compounds  and very little is known about the local impurity effect, mainly because of  the difficulty in obtaining a fresh atomically flat surface by cleaving the single crystal \cite{Ernst11,Aynajian12,Zhou13,Allan13,Schmidt10,Aynajian10,PMostafa16,Kim17}.  Recently, the state-of-the-art molecular beam epitaxy (MBE) technique was developed to fabricate high quality thin films and superlattices comprising of heavy fermion compounds \cite{Shishido10,Shimozawa16}.  In this technique, the crystal growth can be controlled at atomic layer level, as revealed by the superconductivity in CeCoIn$_5$/YbCoIn$_5$ superlattices which consists of  one-unit-cell-thick CeCoIn$_5$ \cite{Mizukami11,Goh12,Yamanaka15}.   Therefore this technique opens up the possibility to produce wide fresh atomically flat terraces, as required for STM measurements. 
Another significant advantage is that MBE is suitable for the preparation of systems with homogeneous  impurity concentration, because the epitaxial growth occurs in non-equilibrium condition at temperatures much lower than the temperature used for single crystal growth by flux method \cite{Shimozawa12}. 
  This is particularly important in heavy fermion compounds, in which several different types of orders often compete and coexist.   Furthermore, owing to the ability to evaporate several atoms simultaneously, the impurity concentration can be controlled very precisely.
Recently, advances in thin-film growth technology combined with STM measurements have provided a unique opportunity to explore novel phenomena in low-dimensional systems with unprecedented control. For example, the {\it in situ} STM observation of epitaxially grown thin films of topological materials and  iron-chalcogenide superconductors provide several unique information on the electronic properties, which cannot be obtained in bulk single crystals \cite{Zhang09,Cheng10,Wang17}. However, such measurements have never been attempted in heavy fermion systems, because of the experimental difficulty for growing their thin films.

\section{Methods}

The STM experiments have been performed with a combined system of MBE and low-temperature ultrahigh vacuum (LT-UHV) STM. The STM head can be cooled down to 300\,mK using Helium-3 refrigerator. Magnetic fields perpendicular to the sample surface can be applied up to 11\,T. All data are obtained with a PtIr tip. All conductance spectra are measured by using a lock-in technique with a modulation voltage $V_{\rm{MOD}}$ and a modulation frequency $f_{\rm{MOD}}$. The $c$ axis oriented epitaxial non-doped and Zn-doped CeCoIn$_5$ films are grown by MBE. The (001) surface of MgF$_2$ with rutile structure ($a$ = 0.462\,nm, $c$ = 0.305\,nm) is used as a substrate. The substrate temperature is kept at 370\,$^{\circ}$C. Each metal element is evaporated from an individually controlled Knudsen cell. The typical deposition rate is 0.01-0.02 nm/s. The area and thickness of the films are 5.0 $\times$ 5.0\,mm$^2$ and 120\,nm, respectively. The quality of films are checked by X-ray diffraction and atomic force microscopy. The films are transfered to the STM head, while being kept in ultrahigh vacuum ($\sim 10^{-8}$ Pa). 

For the theoretical calculations, we use a periodic Anderson model on a square lattice with two-atomic basis. The first atom corresponds to the Ce-atom including conduction- and $f$-electrons. The second atom corresponds to In, including only conduction electrons. In addition to a large electron hopping between the conduction electrons of Ce and In, we include a small hybridization between the $f$-electrons of Ce and the conduction electrons of In. Due to symmetry, we neglect a local hybridization on the Ce atoms. Furthermore, we take into account a small direct transfer between the conduction electrons of different In-atoms. To describe the Kondo effect as necessary for heavy fermion materials, the model contains a local density-density interaction between the $f$-electrons of Ce. We tune the parameters so that a hybridization gap opens at the Fermi energy with a gap-width of $\Delta=50$\,K which corresponds to that of CeCoIn$_5$. The effect of a Zn(1) impurity is included by changing the local energy and  thus reducing the filling of a single atom at an In(1) position by one electron. This model can be regarded as a toy model describing a Zn-impurity in a single layer of CeCoIn$_5$. This model is solved via the real-space dynamical mean field theory (DMFT) which is an extension of DMFT \cite{RevModPhys.68.13} for inhomogeneous systems. Each atom of a finite lattice is mapped onto a separate quantum impurity model and solved self-consistently. Thus, real-space DMFT can treat inhomogeneous systems including impurities. The quantum impurity models are solved via the numerical renormalization group (NRG)\cite{ RevModPhys.80.395}. By using real-space DMFT in combination with NRG, we calculate the LDOS close to a Zn(1) impurity and compare it to an In(1) atom.


\section{Results}

The $c$ axis oriented non-doped CeCoIn$_5$ thin films were grown by MBE in ultrahigh vacuum ($\sim 10^{-8}$\,Pa) on MgF$_2$ substrate. The films are transfered to the STM head, while being kept in ultrahigh vacuum. The superconducting transition temperature $T_c$ of this thin film is 2.0\,K, which is slightly reduced from the single crystal ($T_c$ = 2.3\,K).    The residual resistivity of this film is $\rho_0 \approx  5\,\mu \Omega$cm,  which is comparable to that of high-quality single crystal ($\rho_0\sim 3\,\mu \Omega$cm).  The resistivity shows $T$-linear dependence above $T_c$, which is a hallmark of non-Fermi liquid \cite{Nakajima07}.  This $T$-dependence, along with the residual resistivity value,  is consistent with those of bulk single crystals. The nuclear quadrupole resonance  spectra and NMR relaxation rate of thin films are essentially unchanged from bulk single crystals \cite{Yamanaka15}.  The slight reduction of $T_c$ is likely to be due to the strain effect from the substrate. We note that the $x-T$ phase diagram of impurity substituted  (Ce$_{1-x}$Yb$_x$)CoIn$_5$  thin films \cite{Shimozawa12} is very similar to that of single crystals \cite{Jang14}.  Based on these results, we conclude that the physical properties of  thin films are essentially same as those of bulk single crystals.

 Figure\,1(b) shows a typical topographic image of a non-doped CeCoIn$_5$ thin film.   The cross-sectional profile along the blue line in Fig.\,1(b) is depicted in Fig.\,1(c).   Step structures are clearly observed and the step height is found to be 0.761 nm, which coincides with the $c$ axis lattice constant.  We note that the atomically flat terraces  are much wider than those of cleaved planes of single crystals.  Figure\,1(d) depicts the STM topograph with atomic resolution of non-doped CeCoIn$_5$ thin film.   The spatially periodic bright spots forming the square lattice represent In(1) atoms in CeIn(1) plane.   In fact, the distance between these spots determined by this image is 0.467\,nm, which coincides with the In(1)-In(1) distance.   The randomly distributed large bright spots appear to be the defects of In atoms.  The top surface of the thin films always exhibits the CeIn(1) plane.  We note that single crystals exhibit  three different cleaved surfaces, CeIn(1), In(2) and Co planes \cite{Aynajian12}.  In contrast to In atoms, Ce atoms are not resolved clearly. This is due to the fact that the $p$-orbitals of In atom are extended perpendicular to the surface, while all orbitals of Ce atom are less extended.

\begin{figure}[t]
	\begin{center}
		\includegraphics[width=1.0\linewidth]{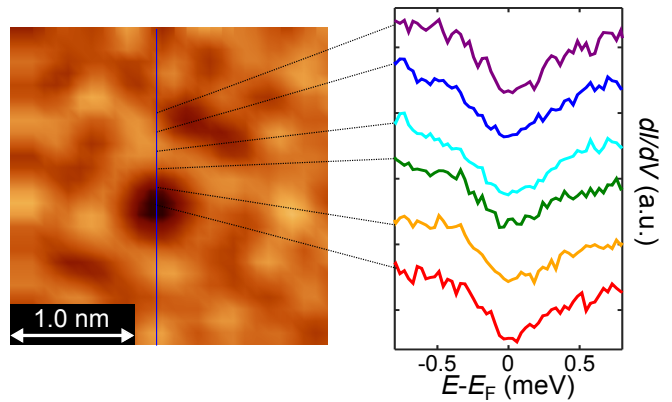}
	\end{center}
	\caption{STM topograph and $dI/dV$ spectra taken in CeCo(In$_{0.993}$Zn$_{0.007}$)$_5$ at 0.5\,K. Tunnel parameters: $V_{\rm{S}}$ = 25\,mV, $I_{\rm{T}}$ = 3.0\,nA, $V_{\rm{MOD}}$ = 30\,$\mu$V, and $f_{\rm{MOD}}$ = 1\,kHz.}
\end{figure}

 Figure\,1(e) shows the tunneling conductance $dI/dV$ spectra, which are proportional to the local density of states (LDOS),   taken at the In(1) site of non-doped CeCoIn$_5$ thin film at 0.5\,K.    The obtained spectrum is essentially the same as the reported one in single crystals\cite{Zhou13,Allan13,Kim17}.   The reduction of LDOS is observed in a wide energy range around the Fermi energy $E_F$.     This reduction arises from the formation of a hybridization gap  $\Delta_{\rm{HG}}$.   At high temperature $T$,  the 4$f$ states are well localized, and only non-$f$ ($s$, $p$, and $d$) dispersive conduction bands are present near $E_{\rm F}$.    With decreasing $T$,   the 4$f$-electrons begin to hybridize with conduction electrons via the Kondo effect and the flat quasiparticle band with 4$f$ character appears near $E_{\rm F}$.  This quasiparticle band  is identified as the Kondo resonance peak in the density of states around the Fermi energy. Besides this flat quasiparticle band, the hybridization opens a gap, which begins to emerge below the coherent temperature $T_{\rm K}$ ($\sim 50$\,K for CeCoIn$_5$).  
In addition to this hybridization gap, a superconducting gap $\Delta_{\rm SC}$ appears at $E_{\rm F}$.   Similar to single crystals, $dI/dV$ remains large  at $E_{\rm F}$ even at 0.5\,K well  below $T_c$, indicating a large residual quasiparticles weight within the superconducting condensate, which is consistent with the results of single crystal.

We note that the local electronic structure is measured mainly through the $p_z$ orbital of In(1) atoms which is well extended perpendicular to the surface. Therefore, as shown in Fig.\,1(f), the spectra do not depend on the tip-position in undoped CeCoIn$_5$. In Zn-doped CeCoIn$_5$, the LDOS at a Zn(1) impurity site is obtained via the direct tunneling process from a Zn impurity (Fig.\,1(g)), while the local electronic structure at a Zn(2) impurity site is obtained indirectly through the In(1) atoms above Zn(2) (Fig.\,1(h)). We denote these spectra as Zn(1) and Zn(2) spectrum.
 
   Figure\,1(i) depicts the topographic image of CeCo(In$_{0.993}$Zn$_{0.007}$)$_5$ thin film. The transition temperature is $T_{\rm c}$ = 1.8\,K, which is slightly reduced by small Zn doping. The dark spots, which are absent in non-doped CeCoIn$_5$, can be identified as  Zn atoms.  The ratio of the numbers of Zn to the numbers of In atoms determined by STM well coincides with the evaporation rate of both elements.    As shown in Fig.\,1(j) (boxed area in Fig.\,1(i)),  in addition to Zn(1) atom residing on the In(1) site, Zn(2) atoms on the In(2) site out of the CeIn(1) plane in the subsurface are observed in-between the In(1) atoms.  The proportion of the Zn atom on the In(1) site is approximately 98\,\% (in comparison with 20\,\% expected for random occupation), indicating a strong preference to occupy the In(1) atom.  This proportion is larger than the substitution of Cd and Hg, ranging from $\sim40$\,\% to $\sim 70$\,\% \cite{Booth09, Daniel05}. Thus,  in the present CeCo(In$_{0.993}$Zn$_{0.007}$)$_5$, nearly 3\,\% of In(1) site is replaced by Zn(1) and less than 0.02\,\% of In(2) is by Zn(2).

\begin{figure}[t]
	\begin{center}
		\includegraphics[width=0.5\linewidth]{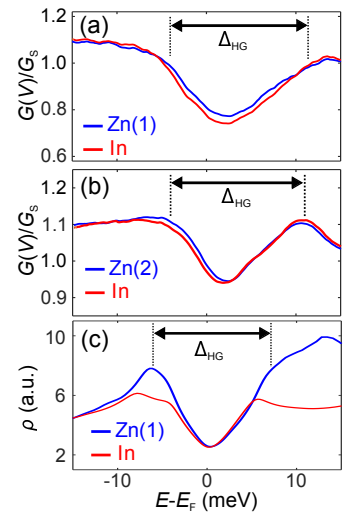}
	\end{center}
	\caption{(a) Tunneling conductance at the In site and Zn(1) site of CeCo(In$_{0.993}$Zn$_{0.007}$)$_5$ at 2.0\,K. Tunnel parameters: $V_{\rm{S}}$ = 25\,mV, $I_{\rm{T}}$ = 5.0\,nA, $V_{\rm{MOD}}$ = 150\,$\mu$V, and $f_{\rm{MOD}}$ = 1\,kHz. (b) Tunneling conductance at the In(2) site and Zn(2) site of CeCo(In$_{0.993}$Zn$_{0.007}$)$_5$ at 2.0\,K. Tunnel parameters: $V_S$ = 25\,mV, $I_T$ = 5.0\,nA, $V_{\rm{MOD}}$ = 150\,$\mu$V, and $f_{\rm{MOD}}$ = 1\,kHz. (c) Calculated local density of states of conduction electrons $\rho$ at Zn(1) site and In site.
}
\end{figure}

The observation of superconducting and hybridization gaps at and around Zn impurities provide direct information of the local electronic structure.  Figure 2 shows the superconducting gap at and around the Zn(1) site in CeCo(In$_{0.993}$Zn$_{0.007}$)$_5$. Remarkably, the superconducting gap spectra around Zn(1) site are essentially unchanged from those around neighboring In sites. 
Figures\,3(a) and (b) compare the conductance spectra of Zn(1) to In sites and the spectra of Zn(2) to In sites, respectively. Here, spectra at In sites are taken at In atoms far away (at least seven lattice site) from any Zn impurity. We note that these spectra are essentially same as those of non-doped CeCoIn5. To compare the intensity of the spectra within the hybridization gap, the conductance, $G$($V$) is normalized by a setpoint conductance, $G_{\rm{S}}$. These data demonstrate that the conductance at both Zn(1) and Zn(2) are essentially same as those at In sites.

\begin{figure*}[t]
	\begin{center}
		\includegraphics[width=1\linewidth]{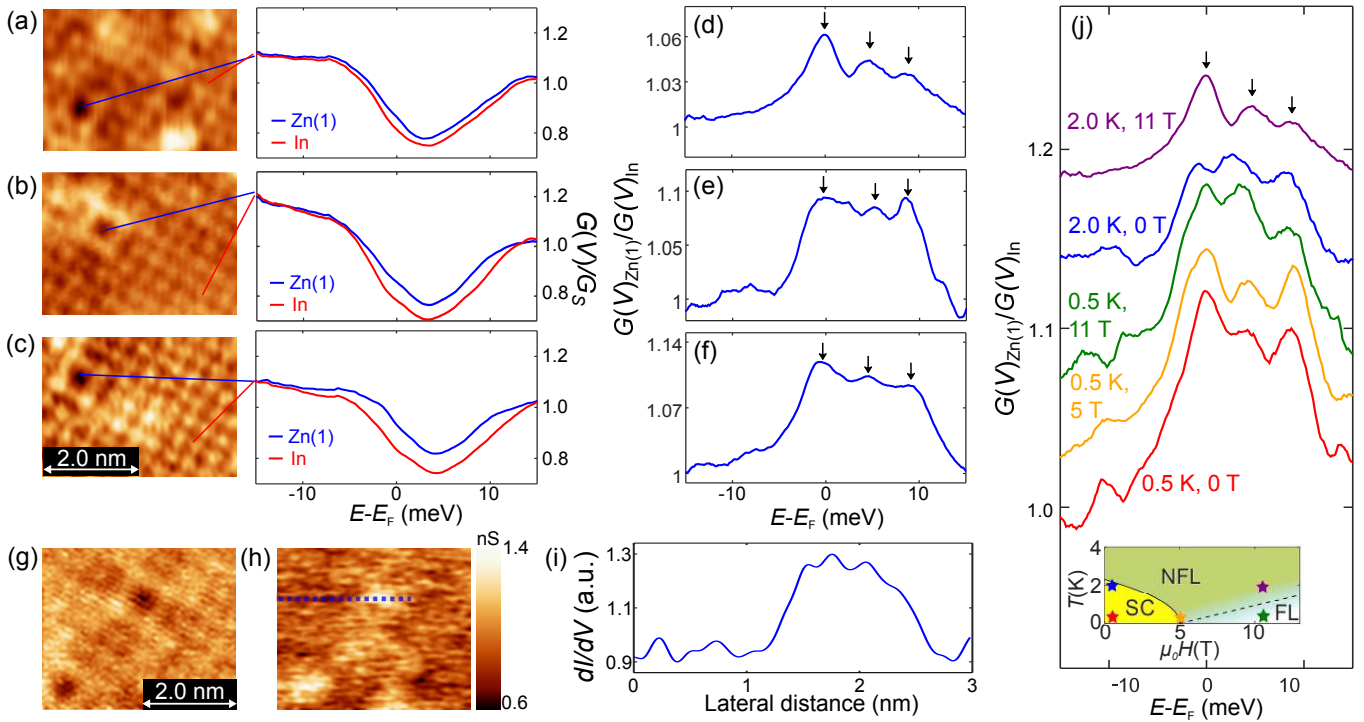}
	\end{center}
	\caption{(a)-(c) Conductance spectra at Zn(1) and In sites in each topographic image in CeCo(In$_{0.993}$Zn$_{0.007}$)$_5$ taken on different terraces at $T$ = 2.0\,K and $B$ = 11\,T. Black spot represents Zn(1) atom. Tunnel parameters: $V_{\rm{S}}$ = 25\,mV, $I_{\rm{T}}$ = 5.0\,nA, $V_{\rm{MOD}}$ = 150\,${\mu}$V, and $f_{\rm{MOD}}$ = 1\,kHz. (d)-(f) Conductance spectra at Zn(1) sites normalized by those at In sites. (g)(h) Topographic image and the corresponding $dI/dV$ image at 0.5\,meV, respectively. These images are taken at $T$ = 0.5\,K and $B$ = 11\,T. (i) The line section taken along blue dotted line in (h). (j) Normalized conductance spectra taken at various temperatures and magnetic fields.  The inset shows the measured points in the $T$-$H$ phase diagram.  SC, NFL, and FL represent superconducting, non-Fermi liquid, and Fermi liquid regimes, respectively. }
\end{figure*}

The absence of any discernible change of the superconducting gap spectra around Zn(1) site immediately indicates that non-superconducting regions are not formed by the Zn impurity, as this would lead to strong suppression of the superconducting gap.  In addition,   in $d$-wave superconductors with sign changing order parameter, a non-superconducting regime creates the Andreev bound states, which are not observed in the present measurements.  Let us here consider two scenarios about the effect of an AFM droplet on the conduction electrons. First, the AFM droplets are fluctuating, which may occur in the dilute case. In this situation, the formation of a magnetic moment reduces the $c$-$f$ Kondo mixing, which changes the hybridization gap and LDOS.  Therefore, it can be observed in our STM measurements. Moreover, the modification of the LDOS affects the superfluid density, which leads to a change of the superconducting gap. Second, if the AFM droplets form a static order, the LDOS around the impurity is largely influenced because the $c$-$f$ Kondo mixing is suppressed similar to the fluctuating case. The influence on the superconductivity is even more dramatic. In the majority of models, static magnetic order competes with $d$-wave superconductivity. Therefore, the superconductivity is strongly suppressed, if static AFM order is induced.  Since our measurements demonstrate that the hybridization and superconducting gaps are almost unchanged, it is unlikely that AFM droplets are induced at and around a Zn impurity.

The absence of changes of the hybridization gap around the nonmagnetic Zn impurity is supported by the calculations based on dynamical mean field theory \cite{RevModPhys.68.13}, as displayed in Fig.\,3(c).  In particular, the gap width is unchanged, which agrees with the experimental results. We use a periodic Anderson model on a square lattice with a two-atomic basis, which takes Ce- and In(1)-atoms into account. More details on the calculations can be found below in the methods section. To simulate the effect of a Zn(1) impurity, we reduce the electron number of a single In(1) atom by one electron and calculate the LDOS around this atom. These calculations are done in the paramagnetic state. 
Although the conduction electron band is disturbed around the Zn(1) impurity, the hybridization between the conduction $s$-, $p$-, and $d$-electrons with the Ce $f$-electrons leads to a gap at the Fermi energy. Our calculations suggest that the width and the shape of this gap is unaltered by including a single impurity. Modifications can only be observed for energies outside the hybridization gap.
It should be stressed that these results, which qualitatively agree with the experimental LDOS around a nonmagnetic Zn impurity within the hybridization gap, were obtained without any induced antiferromagnetic moment. Thus, the present calculations do not provide any evidence for the formation of  AFM droplets around a single nonmagnetic Zn impurity. 

The NMR measurements in Cd-doped CeCoIn$_5$ reported that the induced AFM moment is strongly enhanced at Cd(2) site \cite{Urbano07,Sakai15}.   However,  our observation of little  change of the LDOS both inside and outside the hybridization gap at Zn(2) site indicates that the influence of a nonmagnetic Zn impurity on the electronic structure is much weaker, i.e. Zn(2) atom acts as a very weak scatterer.  In addition, STM topograph shows that the number of Zn(2) impurities is less than 1\,\% of Zn(1) impurities. This indicates that in CeCo(In$_{0.98}$Zn$_{0.02}$)$_5$ in which long range AFM ordering is observed \cite{Yokoyama15}, only 0.06\,\% of In(2) atoms are replaced by Zn atoms. Therefore, it is highly unlikely that the Zn(2) site plays an essential role for the AFM ordering.

Although Zn substitution leads to no discernible change of the LDOS within the hybridization gap at the first glance as shown in Fig.\,4(a)-(c), a tiny modification  still appears to be present.  Figures\,4(d)-(f) display the tunneling conductances at Zn(1) sites on three different terraces normalized by those at In sites locating close to each Zn(1) site. A triple peak structure is clearly visible in the data. This structure is intrinsic and can be observed even when measured with different STM tips. To analyze the spatial dependence of this peak, we show a topographic image around a Zn(1) site in Fig.\,4(g) and corresponding $dI/dV$ mapping of the left peak position in Fig.\,4(h). Obviously, the peak structure can be seen only around the Zn(1) sites; the peak structure spatially spreads out to the neighboring In atoms as shown in the line profile in Fig.\,4(i). We note that the difference at the Zn site in Fig. 4(h) and (i) is emphasized compared to Fig. 4(d)-(f) because of low setpoint bias voltage. We point out that this peak structure cannot be related to any possible induced magnetism around the Zn site.  In fact, as shown in Fig.\,4(j), this structure appears  in a wide region of the $T$-$H$ phase diagram, regardless of  Fermi or non-Fermi liquid states, indicating that the AFM correlations are irrelevant to the peak structure.  This tiny peak structure,  which is not captured by the periodic Anderson model, is an open question, which deserves detailed future study.

\section{Discussion}

Our STM measurements do not provide evidence that AFM droplets are induced around nonmagnetic Zn impurities.  There are two scenarios explaining our measurements. First, AFM droplets are not formed around non-magnetic impurities and the appearance of long range AFM order at higher Zn concentrations is due to the shift of the chemical potential by hole doping.  Second, the induced AFM moments are so tiny that they have no influence on the superconducting and hybridization gaps.
We note that although the phase diagrams of CeCo(In$_{1-x}R_x$)$_5$ for $R$ = Zn, Cd, and Hg,  in which one hole is doped, bear strong resemblance, we cannot rule out the possibility that the electronic structure around an impurity depends on the dopant; While AFM droplets are induced around Cd atoms, they are not around Zn atoms.  In this case, the mechanism of the induction of the static AFM ordering for Zn doping is different from Cd case.

Our results furthermore demonstrate a significant difference of the impact of nonmagnetic impurities between cuprates and layered heavy fermion systems, despite many remarkable common features in both systems.  This difference may originate from the different ground states of cuprates and heavy fermions, which are theoretically described by a Hubbard and a periodic Anderson models, respectively. This might also be related to the fact that the mother material, from which superconductivity emerges when the AFM order is suppressed, is insulating for cuprates, while it is metallic for heavy fermions.

\section{Conclusion}

In summary,  to investigate the local electronic structure around a nonmagnetic Zn impurity in a heavy fermion superconductor in the vicinity of an  AFM QCP, we performed {\it in situ}  STM measurements on CeCo(In$_{1-x}$Zn$_x$)$_5$ thin film grown by MBE.  The hybridization and superconducting gaps are clearly resolved around the Fermi level.   
 What is remarkable is that the superconducting gap at the Zn(1) site in the Ce plane exhibit no discernible change from that at In sites. Moreover, the LDOS within the hybridization gap also exhibits only little change both at Zn(1) and Zn(2) sites.   Thus, our results do not provide any evidence for induced AFM moments around a single Zn impurity.  In addition, Zn(2) atoms outside the Ce plane do not play an important role for the emergence of AFM order and suppression of $T_c$.  The different response to nonmagnetic impurities between cuprates and CeCoIn$_5$ may also provide important information on the coexistence and competition of unconventional superconductivity and antiferromagnetism.  Moreover,  combined MBE-STM system should pave a new way for further experimental studies of exotic electronic properties of heavy fermion compounds.

\begin{acknowledgments}
We thank T. Hanaguri, Y. Hasegawa, K. Ishida, H. Kontani, H. Sakai, Y. Yanase, M. Yokoyama, and Y. Yoshida,   for fruitful discussion. This work was supported by Grants-in-Aid for Scientific Research (KAKENHI) (No.\,25220710, No.\,15H02106, No.\,15H03688, No.\,15H05457), Grants-in-Aid for Scientific Research on Innovative Areas ``Topological Materials Science'' (Nos.\,15H05852 and 15K21717) from Japan Society for the Promotion of Science (JSPS).
\end{acknowledgments}


\bibliography{STM115_PRB}

\providecommand{\noopsort}[1]{}\providecommand{\singleletter}[1]{#1}%
\begin{thebibliography}{48}%
\makeatletter
\providecommand \@ifxundefined [1]{%
 \@ifx{#1\undefined}
}%
\providecommand \@ifnum [1]{%
 \ifnum #1\expandafter \@firstoftwo
 \else \expandafter \@secondoftwo
 \fi
}%
\providecommand \@ifx [1]{%
 \ifx #1\expandafter \@firstoftwo
 \else \expandafter \@secondoftwo
 \fi
}%
\providecommand \natexlab [1]{#1}%
\providecommand \enquote  [1]{``#1''}%
\providecommand \bibnamefont  [1]{#1}%
\providecommand \bibfnamefont [1]{#1}%
\providecommand \citenamefont [1]{#1}%
\providecommand \href@noop [0]{\@secondoftwo}%
\providecommand \href [0]{\begingroup \@sanitize@url \@href}%
\providecommand \@href[1]{\@@startlink{#1}\@@href}%
\providecommand \@@href[1]{\endgroup#1\@@endlink}%
\providecommand \@sanitize@url [0]{\catcode `\\12\catcode `\$12\catcode
  `\&12\catcode `\#12\catcode `\^12\catcode `\_12\catcode `\%12\relax}%
\providecommand \@@startlink[1]{}%
\providecommand \@@endlink[0]{}%
\providecommand \url  [0]{\begingroup\@sanitize@url \@url }%
\providecommand \@url [1]{\endgroup\@href {#1}{\urlprefix }}%
\providecommand \urlprefix  [0]{URL }%
\providecommand \Eprint [0]{\href }%
\providecommand \doibase [0]{http://dx.doi.org/}%
\providecommand \selectlanguage [0]{\@gobble}%
\providecommand \bibinfo  [0]{\@secondoftwo}%
\providecommand \bibfield  [0]{\@secondoftwo}%
\providecommand \translation [1]{[#1]}%
\providecommand \BibitemOpen [0]{}%
\providecommand \bibitemStop [0]{}%
\providecommand \bibitemNoStop [0]{.\EOS\space}%
\providecommand \EOS [0]{\spacefactor3000\relax}%
\providecommand \BibitemShut  [1]{\csname bibitem#1\endcsname}%
\let\auto@bib@innerbib\@empty
\bibitem [{\citenamefont {Mathur}\ \emph {et~al.}(1998)\citenamefont {Mathur},
  \citenamefont {Grosche}, \citenamefont {Julian}, \citenamefont {Walker},
  \citenamefont {Freye}, \citenamefont {Haselwimmer},\ and\ \citenamefont
  {Lonzarich}}]{Mathur98}%
  \BibitemOpen
  \bibfield  {author} {\bibinfo {author} {\bibfnamefont {N.~D.}\ \bibnamefont
  {Mathur}}, \bibinfo {author} {\bibfnamefont {F.~M.}\ \bibnamefont {Grosche}},
  \bibinfo {author} {\bibfnamefont {S.~R.}\ \bibnamefont {Julian}}, \bibinfo
  {author} {\bibfnamefont {I.~R.}\ \bibnamefont {Walker}}, \bibinfo {author}
  {\bibfnamefont {D.~M.}\ \bibnamefont {Freye}}, \bibinfo {author}
  {\bibfnamefont {R.~K.~W.}\ \bibnamefont {Haselwimmer}}, \ and\ \bibinfo
  {author} {\bibfnamefont {G.~G.}\ \bibnamefont {Lonzarich}},\ }\href {\doibase
  10.1038/27838} {\bibfield  {journal} {\bibinfo  {journal} {Nature}\ }\textbf
  {\bibinfo {volume} {394}},\ \bibinfo {pages} {39} (\bibinfo {year}
  {1998})}\BibitemShut {NoStop}%
\bibitem [{\citenamefont {Shibauchi}\ \emph {et~al.}(2014)\citenamefont
  {Shibauchi}, \citenamefont {Carrington},\ and\ \citenamefont
  {Matsuda}}]{Shibauchi13}%
  \BibitemOpen
  \bibfield  {author} {\bibinfo {author} {\bibfnamefont {T.}~\bibnamefont
  {Shibauchi}}, \bibinfo {author} {\bibfnamefont {A.}~\bibnamefont
  {Carrington}}, \ and\ \bibinfo {author} {\bibfnamefont {Y.}~\bibnamefont
  {Matsuda}},\ }\href {\doibase 10.1146/annurev-conmatphys-031113-133921}
  {\bibfield  {journal} {\bibinfo  {journal} {Annu. Rev. Condens. Matter
  Phys.}\ }\textbf {\bibinfo {volume} {5}},\ \bibinfo {pages} {113} (\bibinfo
  {year} {2014})}\BibitemShut {NoStop}%
\bibitem [{\citenamefont {Keimer}\ \emph {et~al.}(2015)\citenamefont {Keimer},
  \citenamefont {Kivelson}, \citenamefont {Norman}, \citenamefont {Uchida},\
  and\ \citenamefont {Zaanen}}]{Keimer15}%
  \BibitemOpen
  \bibfield  {author} {\bibinfo {author} {\bibfnamefont {B.}~\bibnamefont
  {Keimer}}, \bibinfo {author} {\bibfnamefont {S.~A.}\ \bibnamefont
  {Kivelson}}, \bibinfo {author} {\bibfnamefont {M.~R.}\ \bibnamefont
  {Norman}}, \bibinfo {author} {\bibfnamefont {S.}~\bibnamefont {Uchida}}, \
  and\ \bibinfo {author} {\bibfnamefont {J.}~\bibnamefont {Zaanen}},\ }\href
  {\doibase 10.1038/nature14165} {\bibfield  {journal} {\bibinfo  {journal}
  {Nature}\ }\textbf {\bibinfo {volume} {518}},\ \bibinfo {pages} {179}
  (\bibinfo {year} {2015})}\BibitemShut {NoStop}%
\bibitem [{\citenamefont {Balatsky}\ \emph {et~al.}(2006)\citenamefont
  {Balatsky}, \citenamefont {Vekhter},\ and\ \citenamefont {Zhu}}]{Balatsky06}%
  \BibitemOpen
  \bibfield  {author} {\bibinfo {author} {\bibfnamefont {A.~V.}\ \bibnamefont
  {Balatsky}}, \bibinfo {author} {\bibfnamefont {I.}~\bibnamefont {Vekhter}}, \
  and\ \bibinfo {author} {\bibfnamefont {J.~X.}\ \bibnamefont {Zhu}},\ }\href
  {\doibase 10.1103/RevModPhys.78.373} {\bibfield  {journal} {\bibinfo
  {journal} {Rev. Mod. Phys.}\ }\textbf {\bibinfo {volume} {78}},\ \bibinfo
  {pages} {373} (\bibinfo {year} {2006})}\BibitemShut {NoStop}%
\bibitem [{\citenamefont {Alloul}\ \emph {et~al.}(2009)\citenamefont {Alloul},
  \citenamefont {Bobroff}, \citenamefont {Gabay},\ and\ \citenamefont
  {Hirschfeld}}]{Alloul09}%
  \BibitemOpen
  \bibfield  {author} {\bibinfo {author} {\bibfnamefont {H.}~\bibnamefont
  {Alloul}}, \bibinfo {author} {\bibfnamefont {J.}~\bibnamefont {Bobroff}},
  \bibinfo {author} {\bibfnamefont {M.}~\bibnamefont {Gabay}}, \ and\ \bibinfo
  {author} {\bibfnamefont {P.~J.}\ \bibnamefont {Hirschfeld}},\ }\href@noop {}
  {\bibfield  {journal} {\bibinfo  {journal} {Rev. Mod Phys.}\ }\textbf
  {\bibinfo {volume} {81}} (\bibinfo {year} {2009})}\BibitemShut {NoStop}%
\bibitem [{\citenamefont {Pan}\ \emph {et~al.}(2000)\citenamefont {Pan},
  \citenamefont {Hudson}, \citenamefont {Lang}, \citenamefont {Eisaki},
  \citenamefont {Uchida},\ and\ \citenamefont {Davis}}]{Pan00}%
  \BibitemOpen
  \bibfield  {author} {\bibinfo {author} {\bibfnamefont {S.}~\bibnamefont
  {Pan}}, \bibinfo {author} {\bibfnamefont {E.~E.}\ \bibnamefont {Hudson}},
  \bibinfo {author} {\bibfnamefont {K.~K.}\ \bibnamefont {Lang}}, \bibinfo
  {author} {\bibfnamefont {H.}~\bibnamefont {Eisaki}}, \bibinfo {author}
  {\bibfnamefont {S.}~\bibnamefont {Uchida}}, \ and\ \bibinfo {author}
  {\bibfnamefont {J.~J.}\ \bibnamefont {Davis}},\ }\href {\doibase
  10.1038/35001534} {\bibfield  {journal} {\bibinfo  {journal} {Nature}\
  }\textbf {\bibinfo {volume} {403}},\ \bibinfo {pages} {746} (\bibinfo {year}
  {2000})}\BibitemShut {NoStop}%
\bibitem [{\citenamefont {Mahajan}\ \emph {et~al.}(1994)\citenamefont
  {Mahajan}, \citenamefont {Alloul}, \citenamefont {Collin},\ and\
  \citenamefont {Marucco}}]{Mahajan94}%
  \BibitemOpen
  \bibfield  {author} {\bibinfo {author} {\bibfnamefont {A.~V.}\ \bibnamefont
  {Mahajan}}, \bibinfo {author} {\bibfnamefont {H.}~\bibnamefont {Alloul}},
  \bibinfo {author} {\bibfnamefont {G.}~\bibnamefont {Collin}}, \ and\ \bibinfo
  {author} {\bibfnamefont {J.~F.}\ \bibnamefont {Marucco}},\ }\href {\doibase
  10.1103/PhysRevLett.72.3100} {\bibfield  {journal} {\bibinfo  {journal}
  {Phys. Rev. Lett.}\ }\textbf {\bibinfo {volume} {72}},\ \bibinfo {pages}
  {3100} (\bibinfo {year} {1994})}\BibitemShut {NoStop}%
\bibitem [{\citenamefont {Julien}\ \emph {et~al.}(2000)\citenamefont {Julien},
  \citenamefont {Feh\'er}, \citenamefont {Horvati\ifmmode~\acute{c}\else
  \'{c}\fi{}}, \citenamefont {Berthier}, \citenamefont {Bakharev},
  \citenamefont {S\'egransan}, \citenamefont {Collin},\ and\ \citenamefont
  {Marucco}}]{Julien00}%
  \BibitemOpen
  \bibfield  {author} {\bibinfo {author} {\bibfnamefont {M.-H.}\ \bibnamefont
  {Julien}}, \bibinfo {author} {\bibfnamefont {T.}~\bibnamefont {Feh\'er}},
  \bibinfo {author} {\bibfnamefont {M.}~\bibnamefont
  {Horvati\ifmmode~\acute{c}\else \'{c}\fi{}}}, \bibinfo {author}
  {\bibfnamefont {C.}~\bibnamefont {Berthier}}, \bibinfo {author}
  {\bibfnamefont {O.~N.}\ \bibnamefont {Bakharev}}, \bibinfo {author}
  {\bibfnamefont {P.}~\bibnamefont {S\'egransan}}, \bibinfo {author}
  {\bibfnamefont {G.}~\bibnamefont {Collin}}, \ and\ \bibinfo {author}
  {\bibfnamefont {J.-F.}\ \bibnamefont {Marucco}},\ }\href {\doibase
  10.1103/PhysRevLett.84.3422} {\bibfield  {journal} {\bibinfo  {journal}
  {Phys. Rev. Lett.}\ }\textbf {\bibinfo {volume} {84}},\ \bibinfo {pages}
  {3422} (\bibinfo {year} {2000})}\BibitemShut {NoStop}%
\bibitem [{\citenamefont {Hudson}\ \emph {et~al.}(2001)\citenamefont {Hudson},
  \citenamefont {Lang}, \citenamefont {Madhavan}, \citenamefont {Pan},
  \citenamefont {Eisaki}, \citenamefont {Uchida},\ and\ \citenamefont
  {Davis}}]{Hudson01}%
  \BibitemOpen
  \bibfield  {author} {\bibinfo {author} {\bibfnamefont {E.~W.}\ \bibnamefont
  {Hudson}}, \bibinfo {author} {\bibfnamefont {K.~M.}\ \bibnamefont {Lang}},
  \bibinfo {author} {\bibfnamefont {V.}~\bibnamefont {Madhavan}}, \bibinfo
  {author} {\bibfnamefont {S.~H.}\ \bibnamefont {Pan}}, \bibinfo {author}
  {\bibfnamefont {H.}~\bibnamefont {Eisaki}}, \bibinfo {author} {\bibfnamefont
  {S.}~\bibnamefont {Uchida}}, \ and\ \bibinfo {author} {\bibfnamefont {J.~C.}\
  \bibnamefont {Davis}},\ }\href {\doibase 10.1038/35082019} {\bibfield
  {journal} {\bibinfo  {journal} {Nature}\ }\textbf {\bibinfo {volume} {411}},\
  \bibinfo {pages} {920} (\bibinfo {year} {2001})}\BibitemShut {NoStop}%
\bibitem [{\citenamefont {Thompson}\ and\ \citenamefont
  {Fisk}(2012)}]{Thompson12}%
  \BibitemOpen
  \bibfield  {author} {\bibinfo {author} {\bibfnamefont {J.~D.}\ \bibnamefont
  {Thompson}}\ and\ \bibinfo {author} {\bibfnamefont {Z.}~\bibnamefont
  {Fisk}},\ }\href {\doibase 10.1143/JPSJ.81.011002} {\bibfield  {journal}
  {\bibinfo  {journal} {J. Phys. Soc. Jpn.}\ }\textbf {\bibinfo {volume}
  {81}},\ \bibinfo {pages} {011002} (\bibinfo {year} {2012})}\BibitemShut
  {NoStop}%
\bibitem [{\citenamefont {Izawa}\ \emph {et~al.}(2001)\citenamefont {Izawa},
  \citenamefont {Yamaguchi}, \citenamefont {Matsuda}, \citenamefont {Shishido},
  \citenamefont {Settai},\ and\ \citenamefont {Onuki}}]{Izawa01}%
  \BibitemOpen
  \bibfield  {author} {\bibinfo {author} {\bibfnamefont {K.}~\bibnamefont
  {Izawa}}, \bibinfo {author} {\bibfnamefont {H.}~\bibnamefont {Yamaguchi}},
  \bibinfo {author} {\bibfnamefont {Y.}~\bibnamefont {Matsuda}}, \bibinfo
  {author} {\bibfnamefont {H.}~\bibnamefont {Shishido}}, \bibinfo {author}
  {\bibfnamefont {R.}~\bibnamefont {Settai}}, \ and\ \bibinfo {author}
  {\bibfnamefont {Y.}~\bibnamefont {Onuki}},\ }\href {\doibase
  10.1103/PhysRevLett.87.057002} {\bibfield  {journal} {\bibinfo  {journal}
  {Phys. Rev. Lett.}\ }\textbf {\bibinfo {volume} {87}},\ \bibinfo {pages}
  {057002} (\bibinfo {year} {2001})}\BibitemShut {NoStop}%
\bibitem [{\citenamefont {Zhou}\ \emph {et~al.}(2013)\citenamefont {Zhou},
  \citenamefont {Misra1}, \citenamefont {Neto}, \citenamefont {Aynajian},
  \citenamefont {Baumbach}, \citenamefont {Thompson}, \citenamefont {Bauer},\
  and\ \citenamefont {Yazdani}}]{Zhou13}%
  \BibitemOpen
  \bibfield  {author} {\bibinfo {author} {\bibfnamefont {B.~B.}\ \bibnamefont
  {Zhou}}, \bibinfo {author} {\bibfnamefont {S.}~\bibnamefont {Misra1}},
  \bibinfo {author} {\bibfnamefont {E.~H. d.~S.}\ \bibnamefont {Neto}},
  \bibinfo {author} {\bibfnamefont {P.}~\bibnamefont {Aynajian}}, \bibinfo
  {author} {\bibfnamefont {R.~E.}\ \bibnamefont {Baumbach}}, \bibinfo {author}
  {\bibfnamefont {J.~D.}\ \bibnamefont {Thompson}}, \bibinfo {author}
  {\bibfnamefont {E.~D.}\ \bibnamefont {Bauer}}, \ and\ \bibinfo {author}
  {\bibfnamefont {A.}~\bibnamefont {Yazdani}},\ }\href {\doibase
  10.1038/nphys2672} {\bibfield  {journal} {\bibinfo  {journal} {Nat. Phys.}\
  }\textbf {\bibinfo {volume} {9}},\ \bibinfo {pages} {474} (\bibinfo {year}
  {2013})}\BibitemShut {NoStop}%
\bibitem [{\citenamefont {Allan}\ \emph {et~al.}(2013)\citenamefont {Allan},
  \citenamefont {Massee}, \citenamefont {Morr}, \citenamefont {{Van Dyke}},
  \citenamefont {Rost}, \citenamefont {Mackenzie}, \citenamefont {Petrovic},\
  and\ \citenamefont {Davis}}]{Allan13}%
  \BibitemOpen
  \bibfield  {author} {\bibinfo {author} {\bibfnamefont {M.~P.}\ \bibnamefont
  {Allan}}, \bibinfo {author} {\bibfnamefont {F.}~\bibnamefont {Massee}},
  \bibinfo {author} {\bibfnamefont {D.~K.}\ \bibnamefont {Morr}}, \bibinfo
  {author} {\bibfnamefont {J.}~\bibnamefont {{Van Dyke}}}, \bibinfo {author}
  {\bibfnamefont {a.~W.}\ \bibnamefont {Rost}}, \bibinfo {author}
  {\bibfnamefont {a.~P.}\ \bibnamefont {Mackenzie}}, \bibinfo {author}
  {\bibfnamefont {C.}~\bibnamefont {Petrovic}}, \ and\ \bibinfo {author}
  {\bibfnamefont {J.~C.}\ \bibnamefont {Davis}},\ }\href {\doibase
  10.1038/nphys2671} {\bibfield  {journal} {\bibinfo  {journal} {Nat. Phys.}\
  }\textbf {\bibinfo {volume} {9}},\ \bibinfo {pages} {14} (\bibinfo {year}
  {2013})}\BibitemShut {NoStop}%
\bibitem [{\citenamefont {Pham}\ \emph {et~al.}(2006)\citenamefont {Pham},
  \citenamefont {Park}, \citenamefont {Maquilon}, \citenamefont {Thompson},\
  and\ \citenamefont {Fisk}}]{Pham06}%
  \BibitemOpen
  \bibfield  {author} {\bibinfo {author} {\bibfnamefont {L.~D.}\ \bibnamefont
  {Pham}}, \bibinfo {author} {\bibfnamefont {T.}~\bibnamefont {Park}}, \bibinfo
  {author} {\bibfnamefont {S.}~\bibnamefont {Maquilon}}, \bibinfo {author}
  {\bibfnamefont {J.~D.}\ \bibnamefont {Thompson}}, \ and\ \bibinfo {author}
  {\bibfnamefont {Z.}~\bibnamefont {Fisk}},\ }\href {\doibase
  10.1103/PhysRevLett.97.056404} {\bibfield  {journal} {\bibinfo  {journal}
  {Phys. Rev. Lett.}\ }\textbf {\bibinfo {volume} {97}},\ \bibinfo {pages}
  {056404} (\bibinfo {year} {2006})}\BibitemShut {NoStop}%
\bibitem [{\citenamefont {Yokoyama}\ \emph {et~al.}(2015)\citenamefont
  {Yokoyama}, \citenamefont {Mashiko}, \citenamefont {Otaka}, \citenamefont
  {Sakon}, \citenamefont {Fujimura}, \citenamefont {Tenya}, \citenamefont
  {Kondo}, \citenamefont {Kindo}, \citenamefont {Ikeda}, \citenamefont
  {Yoshizawa}, \citenamefont {Shimizu}, \citenamefont {Kono},\ and\
  \citenamefont {Sakakibara}}]{Yokoyama15}%
  \BibitemOpen
  \bibfield  {author} {\bibinfo {author} {\bibfnamefont {M.}~\bibnamefont
  {Yokoyama}}, \bibinfo {author} {\bibfnamefont {H.}~\bibnamefont {Mashiko}},
  \bibinfo {author} {\bibfnamefont {R.}~\bibnamefont {Otaka}}, \bibinfo
  {author} {\bibfnamefont {Y.}~\bibnamefont {Sakon}}, \bibinfo {author}
  {\bibfnamefont {K.}~\bibnamefont {Fujimura}}, \bibinfo {author}
  {\bibfnamefont {K.}~\bibnamefont {Tenya}}, \bibinfo {author} {\bibfnamefont
  {A.}~\bibnamefont {Kondo}}, \bibinfo {author} {\bibfnamefont
  {K.}~\bibnamefont {Kindo}}, \bibinfo {author} {\bibfnamefont
  {Y.}~\bibnamefont {Ikeda}}, \bibinfo {author} {\bibfnamefont
  {H.}~\bibnamefont {Yoshizawa}}, \bibinfo {author} {\bibfnamefont
  {Y.}~\bibnamefont {Shimizu}}, \bibinfo {author} {\bibfnamefont
  {Y.}~\bibnamefont {Kono}}, \ and\ \bibinfo {author} {\bibfnamefont
  {T.}~\bibnamefont {Sakakibara}},\ }\href {\doibase
  10.1103/PhysRevB.92.184509} {\bibfield  {journal} {\bibinfo  {journal} {Phys.
  Rev. B}\ }\textbf {\bibinfo {volume} {92}},\ \bibinfo {pages} {184509}
  (\bibinfo {year} {2015})}\BibitemShut {NoStop}%
\bibitem [{\citenamefont {Booth}\ \emph {et~al.}(2009)\citenamefont {Booth},
  \citenamefont {Bauer}, \citenamefont {Bianchi}, \citenamefont {Ronning},
  \citenamefont {Thompson}, \citenamefont {Sarrao}, \citenamefont {Cho},
  \citenamefont {Chan}, \citenamefont {Capan},\ and\ \citenamefont
  {Fisk}}]{Booth09}%
  \BibitemOpen
  \bibfield  {author} {\bibinfo {author} {\bibfnamefont {C.~H.}\ \bibnamefont
  {Booth}}, \bibinfo {author} {\bibfnamefont {E.~D.}\ \bibnamefont {Bauer}},
  \bibinfo {author} {\bibfnamefont {A.~D.}\ \bibnamefont {Bianchi}}, \bibinfo
  {author} {\bibfnamefont {F.}~\bibnamefont {Ronning}}, \bibinfo {author}
  {\bibfnamefont {J.~D.}\ \bibnamefont {Thompson}}, \bibinfo {author}
  {\bibfnamefont {J.~L.}\ \bibnamefont {Sarrao}}, \bibinfo {author}
  {\bibfnamefont {J.~Y.}\ \bibnamefont {Cho}}, \bibinfo {author} {\bibfnamefont
  {J.~Y.}\ \bibnamefont {Chan}}, \bibinfo {author} {\bibfnamefont
  {C.}~\bibnamefont {Capan}}, \ and\ \bibinfo {author} {\bibfnamefont
  {Z.}~\bibnamefont {Fisk}},\ }\href {\doibase 10.1103/PhysRevB.79.144519}
  {\bibfield  {journal} {\bibinfo  {journal} {Phys. Rev. B}\ }\textbf {\bibinfo
  {volume} {79}},\ \bibinfo {pages} {144519} (\bibinfo {year}
  {2009})}\BibitemShut {NoStop}%
\bibitem [{\citenamefont {Gofryk}\ \emph {et~al.}(2012)\citenamefont {Gofryk},
  \citenamefont {Ronning}, \citenamefont {Zhu}, \citenamefont {Ou},
  \citenamefont {Tobash}, \citenamefont {Stoyko}, \citenamefont {Lu},
  \citenamefont {Mar}, \citenamefont {Park}, \citenamefont {Bauer},
  \citenamefont {Thompson},\ and\ \citenamefont {Fisk}}]{Gofryk12}%
  \BibitemOpen
  \bibfield  {author} {\bibinfo {author} {\bibfnamefont {K.}~\bibnamefont
  {Gofryk}}, \bibinfo {author} {\bibfnamefont {F.}~\bibnamefont {Ronning}},
  \bibinfo {author} {\bibfnamefont {J.-X.}\ \bibnamefont {Zhu}}, \bibinfo
  {author} {\bibfnamefont {M.~N.}\ \bibnamefont {Ou}}, \bibinfo {author}
  {\bibfnamefont {P.~H.}\ \bibnamefont {Tobash}}, \bibinfo {author}
  {\bibfnamefont {S.~S.}\ \bibnamefont {Stoyko}}, \bibinfo {author}
  {\bibfnamefont {X.}~\bibnamefont {Lu}}, \bibinfo {author} {\bibfnamefont
  {A.}~\bibnamefont {Mar}}, \bibinfo {author} {\bibfnamefont {T.}~\bibnamefont
  {Park}}, \bibinfo {author} {\bibfnamefont {E.~D.}\ \bibnamefont {Bauer}},
  \bibinfo {author} {\bibfnamefont {J.~D.}\ \bibnamefont {Thompson}}, \ and\
  \bibinfo {author} {\bibfnamefont {Z.}~\bibnamefont {Fisk}},\ }\href {\doibase
  10.1103/PhysRevLett.109.186402} {\bibfield  {journal} {\bibinfo  {journal}
  {Phys. Rev. Lett.}\ }\textbf {\bibinfo {volume} {109}},\ \bibinfo {pages}
  {186402} (\bibinfo {year} {2012})}\BibitemShut {NoStop}%
\bibitem [{\citenamefont {Bauer}\ \emph {et~al.}(2006)\citenamefont {Bauer},
  \citenamefont {Ronning}, \citenamefont {Capan}, \citenamefont {Graf},
  \citenamefont {Vandervelde}, \citenamefont {Yuan}, \citenamefont {Salamon},
  \citenamefont {Mixson}, \citenamefont {Moreno}, \citenamefont {Brown},
  \citenamefont {Thompson}, \citenamefont {Movshovich}, \citenamefont
  {Hundley}, \citenamefont {Sarrao}, \citenamefont {Pagliuso},\ and\
  \citenamefont {Kauzlarich}}]{Bauer06}%
  \BibitemOpen
  \bibfield  {author} {\bibinfo {author} {\bibfnamefont {E.~D.}\ \bibnamefont
  {Bauer}}, \bibinfo {author} {\bibfnamefont {F.}~\bibnamefont {Ronning}},
  \bibinfo {author} {\bibfnamefont {C.}~\bibnamefont {Capan}}, \bibinfo
  {author} {\bibfnamefont {M.~J.}\ \bibnamefont {Graf}}, \bibinfo {author}
  {\bibfnamefont {D.}~\bibnamefont {Vandervelde}}, \bibinfo {author}
  {\bibfnamefont {H.~Q.}\ \bibnamefont {Yuan}}, \bibinfo {author}
  {\bibfnamefont {M.~B.}\ \bibnamefont {Salamon}}, \bibinfo {author}
  {\bibfnamefont {D.~J.}\ \bibnamefont {Mixson}}, \bibinfo {author}
  {\bibfnamefont {N.~O.}\ \bibnamefont {Moreno}}, \bibinfo {author}
  {\bibfnamefont {S.~R.}\ \bibnamefont {Brown}}, \bibinfo {author}
  {\bibfnamefont {J.~D.}\ \bibnamefont {Thompson}}, \bibinfo {author}
  {\bibfnamefont {R.}~\bibnamefont {Movshovich}}, \bibinfo {author}
  {\bibfnamefont {M.~F.}\ \bibnamefont {Hundley}}, \bibinfo {author}
  {\bibfnamefont {J.~L.}\ \bibnamefont {Sarrao}}, \bibinfo {author}
  {\bibfnamefont {P.~G.}\ \bibnamefont {Pagliuso}}, \ and\ \bibinfo {author}
  {\bibfnamefont {S.~M.}\ \bibnamefont {Kauzlarich}},\ }\href {\doibase
  10.1103/PhysRevB.73.245109} {\bibfield  {journal} {\bibinfo  {journal} {Phys.
  Rev. B}\ }\textbf {\bibinfo {volume} {73}},\ \bibinfo {pages} {245109}
  (\bibinfo {year} {2006})}\BibitemShut {NoStop}%
\bibitem [{\citenamefont {Ramos}\ \emph {et~al.}(2010)\citenamefont {Ramos},
  \citenamefont {Fontes}, \citenamefont {Hering}, \citenamefont {Continentino},
  \citenamefont {Baggio-Saitovich}, \citenamefont {Neto}, \citenamefont
  {Bittar}, \citenamefont {Pagliuso}, \citenamefont {Bauer}, \citenamefont
  {Sarrao},\ and\ \citenamefont {Thompson}}]{Romas10}%
  \BibitemOpen
  \bibfield  {author} {\bibinfo {author} {\bibfnamefont {S.~M.}\ \bibnamefont
  {Ramos}}, \bibinfo {author} {\bibfnamefont {M.~B.}\ \bibnamefont {Fontes}},
  \bibinfo {author} {\bibfnamefont {E.~N.}\ \bibnamefont {Hering}}, \bibinfo
  {author} {\bibfnamefont {M.~A.}\ \bibnamefont {Continentino}}, \bibinfo
  {author} {\bibfnamefont {E.}~\bibnamefont {Baggio-Saitovich}}, \bibinfo
  {author} {\bibfnamefont {F.~D.}\ \bibnamefont {Neto}}, \bibinfo {author}
  {\bibfnamefont {E.~M.}\ \bibnamefont {Bittar}}, \bibinfo {author}
  {\bibfnamefont {P.~G.}\ \bibnamefont {Pagliuso}}, \bibinfo {author}
  {\bibfnamefont {E.~D.}\ \bibnamefont {Bauer}}, \bibinfo {author}
  {\bibfnamefont {J.~L.}\ \bibnamefont {Sarrao}}, \ and\ \bibinfo {author}
  {\bibfnamefont {J.~D.}\ \bibnamefont {Thompson}},\ }\href {\doibase
  10.1103/PhysRevLett.105.126401} {\bibfield  {journal} {\bibinfo  {journal}
  {Phys. Rev. Lett.}\ }\textbf {\bibinfo {volume} {105}},\ \bibinfo {pages}
  {126401} (\bibinfo {year} {2010})}\BibitemShut {NoStop}%
\bibitem [{\citenamefont {Paglione}\ \emph {et~al.}(2007)\citenamefont
  {Paglione}, \citenamefont {Sayles}, \citenamefont {Ho}, \citenamefont
  {Jeffries},\ and\ \citenamefont {Maple}}]{Paglione07}%
  \BibitemOpen
  \bibfield  {author} {\bibinfo {author} {\bibfnamefont {J.}~\bibnamefont
  {Paglione}}, \bibinfo {author} {\bibfnamefont {T.~A.}\ \bibnamefont
  {Sayles}}, \bibinfo {author} {\bibfnamefont {P.-C.}\ \bibnamefont {Ho}},
  \bibinfo {author} {\bibfnamefont {J.~R.}\ \bibnamefont {Jeffries}}, \ and\
  \bibinfo {author} {\bibfnamefont {M.~B.}\ \bibnamefont {Maple}},\ }\href
  {\doibase 10.1038/nphys711} {\bibfield  {journal} {\bibinfo  {journal} {Nat.
  Phys.}\ }\textbf {\bibinfo {volume} {3}},\ \bibinfo {pages} {703 } (\bibinfo
  {year} {2007})}\BibitemShut {NoStop}%
\bibitem [{\citenamefont {Petrovic}\ \emph {et~al.}(2002)\citenamefont
  {Petrovic}, \citenamefont {Bud'ko}, \citenamefont {Kogan},\ and\
  \citenamefont {Canfield}}]{Petrovic02}%
  \BibitemOpen
  \bibfield  {author} {\bibinfo {author} {\bibfnamefont {C.}~\bibnamefont
  {Petrovic}}, \bibinfo {author} {\bibfnamefont {S.~L.}\ \bibnamefont
  {Bud'ko}}, \bibinfo {author} {\bibfnamefont {V.~G.}\ \bibnamefont {Kogan}}, \
  and\ \bibinfo {author} {\bibfnamefont {P.~C.}\ \bibnamefont {Canfield}},\
  }\href {\doibase 10.1103/PhysRevB.66.054534} {\bibfield  {journal} {\bibinfo
  {journal} {Phys. Rev. B}\ }\textbf {\bibinfo {volume} {66}},\ \bibinfo
  {pages} {054534} (\bibinfo {year} {2002})}\BibitemShut {NoStop}%
\bibitem [{\citenamefont {Bauer}\ \emph {et~al.}(2011)\citenamefont {Bauer},
  \citenamefont {Yang}, \citenamefont {Capan}, \citenamefont {Urbano},
  \citenamefont {Miclea}, \citenamefont {Sakai}, \citenamefont {Ronning},
  \citenamefont {Graf}, \citenamefont {Balatsky}, \citenamefont {Movshovich},
  \citenamefont {Bianchi}, \citenamefont {Reyes}, \citenamefont {Kuhns},
  \citenamefont {Thompson},\ and\ \citenamefont {Fisk}}]{Bauer11}%
  \BibitemOpen
  \bibfield  {author} {\bibinfo {author} {\bibfnamefont {E.~D.}\ \bibnamefont
  {Bauer}}, \bibinfo {author} {\bibfnamefont {Y.-f.}\ \bibnamefont {Yang}},
  \bibinfo {author} {\bibfnamefont {C.}~\bibnamefont {Capan}}, \bibinfo
  {author} {\bibfnamefont {R.~R.}\ \bibnamefont {Urbano}}, \bibinfo {author}
  {\bibfnamefont {C.~F.}\ \bibnamefont {Miclea}}, \bibinfo {author}
  {\bibfnamefont {H.}~\bibnamefont {Sakai}}, \bibinfo {author} {\bibfnamefont
  {F.}~\bibnamefont {Ronning}}, \bibinfo {author} {\bibfnamefont {M.~J.}\
  \bibnamefont {Graf}}, \bibinfo {author} {\bibfnamefont {A.~V.}\ \bibnamefont
  {Balatsky}}, \bibinfo {author} {\bibfnamefont {R.}~\bibnamefont
  {Movshovich}}, \bibinfo {author} {\bibfnamefont {A.~D.}\ \bibnamefont
  {Bianchi}}, \bibinfo {author} {\bibfnamefont {A.~P.}\ \bibnamefont {Reyes}},
  \bibinfo {author} {\bibfnamefont {P.~L.}\ \bibnamefont {Kuhns}}, \bibinfo
  {author} {\bibfnamefont {J.~D.}\ \bibnamefont {Thompson}}, \ and\ \bibinfo
  {author} {\bibfnamefont {Z.}~\bibnamefont {Fisk}},\ }\href {\doibase
  10.1073/pnas.1103965108} {\bibfield  {journal} {\bibinfo  {journal} {Proc.
  Natl. Acad. Sci.}\ }\textbf {\bibinfo {volume} {108}},\ \bibinfo {pages}
  {6857} (\bibinfo {year} {2011})}\BibitemShut {NoStop}%
\bibitem [{\citenamefont {Castro~Neto}\ and\ \citenamefont
  {Jones}(2000)}]{Neto00}%
  \BibitemOpen
  \bibfield  {author} {\bibinfo {author} {\bibfnamefont {A.~H.}\ \bibnamefont
  {Castro~Neto}}\ and\ \bibinfo {author} {\bibfnamefont {B.~A.}\ \bibnamefont
  {Jones}},\ }\href {\doibase 10.1103/PhysRevB.62.14975} {\bibfield  {journal}
  {\bibinfo  {journal} {Phys. Rev. B}\ }\textbf {\bibinfo {volume} {62}},\
  \bibinfo {pages} {14975} (\bibinfo {year} {2000})}\BibitemShut {NoStop}%
\bibitem [{\citenamefont {Millis}\ \emph {et~al.}(2001)\citenamefont {Millis},
  \citenamefont {Morr},\ and\ \citenamefont {Schmalian}}]{Millis01}%
  \BibitemOpen
  \bibfield  {author} {\bibinfo {author} {\bibfnamefont {A.~J.}\ \bibnamefont
  {Millis}}, \bibinfo {author} {\bibfnamefont {D.~K.}\ \bibnamefont {Morr}}, \
  and\ \bibinfo {author} {\bibfnamefont {J.}~\bibnamefont {Schmalian}},\ }\href
  {\doibase 10.1103/PhysRevLett.87.167202} {\bibfield  {journal} {\bibinfo
  {journal} {Phys. Rev. Lett.}\ }\textbf {\bibinfo {volume} {87}},\ \bibinfo
  {pages} {167202} (\bibinfo {year} {2001})}\BibitemShut {NoStop}%
\bibitem [{\citenamefont {Figgins}\ and\ \citenamefont
  {Morr}(2011)}]{Figgins11}%
  \BibitemOpen
  \bibfield  {author} {\bibinfo {author} {\bibfnamefont {J.}~\bibnamefont
  {Figgins}}\ and\ \bibinfo {author} {\bibfnamefont {D.~K.}\ \bibnamefont
  {Morr}},\ }\href {\doibase 10.1103/PhysRevLett.107.066401} {\bibfield
  {journal} {\bibinfo  {journal} {Phys. Rev. Lett.}\ }\textbf {\bibinfo
  {volume} {107}},\ \bibinfo {pages} {066401} (\bibinfo {year}
  {2011})}\BibitemShut {NoStop}%
\bibitem [{\citenamefont {Urbano}\ \emph {et~al.}(2007)\citenamefont {Urbano},
  \citenamefont {Young}, \citenamefont {Curro}, \citenamefont {Thompson},
  \citenamefont {Pham},\ and\ \citenamefont {Fisk}}]{Urbano07}%
  \BibitemOpen
  \bibfield  {author} {\bibinfo {author} {\bibfnamefont {R.~R.}\ \bibnamefont
  {Urbano}}, \bibinfo {author} {\bibfnamefont {B.-L.}\ \bibnamefont {Young}},
  \bibinfo {author} {\bibfnamefont {N.~J.}\ \bibnamefont {Curro}}, \bibinfo
  {author} {\bibfnamefont {J.~D.}\ \bibnamefont {Thompson}}, \bibinfo {author}
  {\bibfnamefont {L.~D.}\ \bibnamefont {Pham}}, \ and\ \bibinfo {author}
  {\bibfnamefont {Z.}~\bibnamefont {Fisk}},\ }\href {\doibase
  10.1103/PhysRevLett.99.146402} {\bibfield  {journal} {\bibinfo  {journal}
  {Phys. Rev. Lett.}\ }\textbf {\bibinfo {volume} {99}},\ \bibinfo {pages}
  {146402} (\bibinfo {year} {2007})}\BibitemShut {NoStop}%
\bibitem [{\citenamefont {Seo}\ \emph {et~al.}(2014)\citenamefont {Seo},
  \citenamefont {Lu}, \citenamefont {Zhu}, \citenamefont {Urbano},
  \citenamefont {Curro}, \citenamefont {Bauer}, \citenamefont {Sidorov},
  \citenamefont {Pham}, \citenamefont {Park}, \citenamefont {Fisk},\ and\
  \citenamefont {Thompson}}]{Seo14}%
  \BibitemOpen
  \bibfield  {author} {\bibinfo {author} {\bibfnamefont {S.}~\bibnamefont
  {Seo}}, \bibinfo {author} {\bibfnamefont {X.}~\bibnamefont {Lu}}, \bibinfo
  {author} {\bibfnamefont {J.-X.}\ \bibnamefont {Zhu}}, \bibinfo {author}
  {\bibfnamefont {R.~R.}\ \bibnamefont {Urbano}}, \bibinfo {author}
  {\bibfnamefont {N.}~\bibnamefont {Curro}}, \bibinfo {author} {\bibfnamefont
  {E.~D.}\ \bibnamefont {Bauer}}, \bibinfo {author} {\bibfnamefont {V.~A.}\
  \bibnamefont {Sidorov}}, \bibinfo {author} {\bibfnamefont {L.~D.}\
  \bibnamefont {Pham}}, \bibinfo {author} {\bibfnamefont {T.}~\bibnamefont
  {Park}}, \bibinfo {author} {\bibfnamefont {Z.}~\bibnamefont {Fisk}}, \ and\
  \bibinfo {author} {\bibfnamefont {J.~D.}\ \bibnamefont {Thompson}},\ }\href
  {\doibase 10.1038/nphys2820} {\bibfield  {journal} {\bibinfo  {journal} {Nat.
  Phys.}\ }\textbf {\bibinfo {volume} {10}},\ \bibinfo {pages} {120 } (\bibinfo
  {year} {2014})}\BibitemShut {NoStop}%
\bibitem [{\citenamefont {Sakai}\ \emph {et~al.}(2015)\citenamefont {Sakai},
  \citenamefont {Ronning}, \citenamefont {Zhu}, \citenamefont {Wakeham},
  \citenamefont {Yasuoka}, \citenamefont {Tokunaga}, \citenamefont {Kambe},
  \citenamefont {Bauer},\ and\ \citenamefont {Thompson}}]{Sakai15}%
  \BibitemOpen
  \bibfield  {author} {\bibinfo {author} {\bibfnamefont {H.}~\bibnamefont
  {Sakai}}, \bibinfo {author} {\bibfnamefont {F.}~\bibnamefont {Ronning}},
  \bibinfo {author} {\bibfnamefont {J.-X.}\ \bibnamefont {Zhu}}, \bibinfo
  {author} {\bibfnamefont {N.}~\bibnamefont {Wakeham}}, \bibinfo {author}
  {\bibfnamefont {H.}~\bibnamefont {Yasuoka}}, \bibinfo {author} {\bibfnamefont
  {Y.}~\bibnamefont {Tokunaga}}, \bibinfo {author} {\bibfnamefont
  {S.}~\bibnamefont {Kambe}}, \bibinfo {author} {\bibfnamefont {E.~D.}\
  \bibnamefont {Bauer}}, \ and\ \bibinfo {author} {\bibfnamefont {J.~D.}\
  \bibnamefont {Thompson}},\ }\href {\doibase 10.1103/PhysRevB.92.121105}
  {\bibfield  {journal} {\bibinfo  {journal} {Phys. Rev. B}\ }\textbf {\bibinfo
  {volume} {92}},\ \bibinfo {pages} {121105} (\bibinfo {year}
  {2015})}\BibitemShut {NoStop}%
\bibitem [{\citenamefont {Ernst}\ \emph {et~al.}(2011)\citenamefont {Ernst},
  \citenamefont {Kirchner}, \citenamefont {Krellner}, \citenamefont {Geibel},
  \citenamefont {Steglich},\ and\ \citenamefont {Wirth}}]{Ernst11}%
  \BibitemOpen
  \bibfield  {author} {\bibinfo {author} {\bibfnamefont {S.}~\bibnamefont
  {Ernst}}, \bibinfo {author} {\bibfnamefont {S.}~\bibnamefont {Kirchner}},
  \bibinfo {author} {\bibfnamefont {C.}~\bibnamefont {Krellner}}, \bibinfo
  {author} {\bibfnamefont {C.}~\bibnamefont {Geibel}}, \bibinfo {author}
  {\bibfnamefont {F.}~\bibnamefont {Steglich}}, \ and\ \bibinfo {author}
  {\bibfnamefont {S.}~\bibnamefont {Wirth}},\ }\href@noop {} {\bibfield
  {journal} {\bibinfo  {journal} {Nature}\ }\textbf {\bibinfo {volume} {474}},\
  \bibinfo {pages} {362} (\bibinfo {year} {2011})}\BibitemShut {NoStop}%
\bibitem [{\citenamefont {Aynajian}\ \emph {et~al.}(2012)\citenamefont
  {Aynajian}, \citenamefont {{da Silva Neto}}, \citenamefont {Gyenis},
  \citenamefont {Baumbach}, \citenamefont {Thompson}, \citenamefont {Fisk},
  \citenamefont {Bauer},\ and\ \citenamefont {Yazdani}}]{Aynajian12}%
  \BibitemOpen
  \bibfield  {author} {\bibinfo {author} {\bibfnamefont {P.}~\bibnamefont
  {Aynajian}}, \bibinfo {author} {\bibfnamefont {E.~H.}\ \bibnamefont {{da
  Silva Neto}}}, \bibinfo {author} {\bibfnamefont {A.}~\bibnamefont {Gyenis}},
  \bibinfo {author} {\bibfnamefont {R.~E.}\ \bibnamefont {Baumbach}}, \bibinfo
  {author} {\bibfnamefont {J.~D.}\ \bibnamefont {Thompson}}, \bibinfo {author}
  {\bibfnamefont {Z.}~\bibnamefont {Fisk}}, \bibinfo {author} {\bibfnamefont
  {E.~D.}\ \bibnamefont {Bauer}}, \ and\ \bibinfo {author} {\bibfnamefont
  {A.}~\bibnamefont {Yazdani}},\ }\href {\doibase 10.1038/nature11204}
  {\bibfield  {journal} {\bibinfo  {journal} {Nature}\ }\textbf {\bibinfo
  {volume} {486}},\ \bibinfo {pages} {201} (\bibinfo {year}
  {2012})}\BibitemShut {NoStop}%
\bibitem [{\citenamefont {Schmidt}\ \emph {et~al.}(2010)\citenamefont
  {Schmidt}, \citenamefont {Hamidian}, \citenamefont {Wahl}, \citenamefont
  {Meier}, \citenamefont {Balatsky}, \citenamefont {Garrett}, \citenamefont
  {Williams}, \citenamefont {Luke},\ and\ \citenamefont {Davis}}]{Schmidt10}%
  \BibitemOpen
  \bibfield  {author} {\bibinfo {author} {\bibfnamefont {A.~R.}\ \bibnamefont
  {Schmidt}}, \bibinfo {author} {\bibfnamefont {M.~H.}\ \bibnamefont
  {Hamidian}}, \bibinfo {author} {\bibfnamefont {P.}~\bibnamefont {Wahl}},
  \bibinfo {author} {\bibfnamefont {F.}~\bibnamefont {Meier}}, \bibinfo
  {author} {\bibfnamefont {a.~V.}\ \bibnamefont {Balatsky}}, \bibinfo {author}
  {\bibfnamefont {J.~D.}\ \bibnamefont {Garrett}}, \bibinfo {author}
  {\bibfnamefont {T.~J.}\ \bibnamefont {Williams}}, \bibinfo {author}
  {\bibfnamefont {G.~M.}\ \bibnamefont {Luke}}, \ and\ \bibinfo {author}
  {\bibfnamefont {J.~C.~S.}\ \bibnamefont {Davis}},\ }\href {\doibase
  10.1038/nature09073} {\bibfield  {journal} {\bibinfo  {journal} {Nature}\
  }\textbf {\bibinfo {volume} {465}},\ \bibinfo {pages} {570} (\bibinfo {year}
  {2010})}\BibitemShut {NoStop}%
\bibitem [{\citenamefont {Aynajian}\ \emph {et~al.}(2010)\citenamefont
  {Aynajian}, \citenamefont {{da Silva Neto}}, \citenamefont {Parker},
  \citenamefont {Huang}, \citenamefont {Pasupathy}, \citenamefont {Mydosh},\
  and\ \citenamefont {Yazdani}}]{Aynajian10}%
  \BibitemOpen
  \bibfield  {author} {\bibinfo {author} {\bibfnamefont {P.}~\bibnamefont
  {Aynajian}}, \bibinfo {author} {\bibfnamefont {E.~H.}\ \bibnamefont {{da
  Silva Neto}}}, \bibinfo {author} {\bibfnamefont {C.~V.}\ \bibnamefont
  {Parker}}, \bibinfo {author} {\bibfnamefont {Y.}~\bibnamefont {Huang}},
  \bibinfo {author} {\bibfnamefont {A.}~\bibnamefont {Pasupathy}}, \bibinfo
  {author} {\bibfnamefont {J.}~\bibnamefont {Mydosh}}, \ and\ \bibinfo {author}
  {\bibfnamefont {A.}~\bibnamefont {Yazdani}},\ }\href {\doibase
  10.1073/pnas.1005892107} {\bibfield  {journal} {\bibinfo  {journal} {Proc.
  Natl. Acad. Sci.}\ }\textbf {\bibinfo {volume} {107}},\ \bibinfo {pages}
  {10383} (\bibinfo {year} {2010})}\BibitemShut {NoStop}%
\bibitem [{\citenamefont {Enayat}\ \emph {et~al.}(2016)\citenamefont {Enayat},
  \citenamefont {Sun}, \citenamefont {Maldonado}, \citenamefont {Suderow},
  \citenamefont {Seiro}, \citenamefont {Geibel}, \citenamefont {Wirth},
  \citenamefont {Steglich},\ and\ \citenamefont {Wahl}}]{PMostafa16}%
  \BibitemOpen
  \bibfield  {author} {\bibinfo {author} {\bibfnamefont {M.}~\bibnamefont
  {Enayat}}, \bibinfo {author} {\bibfnamefont {Z.}~\bibnamefont {Sun}},
  \bibinfo {author} {\bibfnamefont {A.}~\bibnamefont {Maldonado}}, \bibinfo
  {author} {\bibfnamefont {H.}~\bibnamefont {Suderow}}, \bibinfo {author}
  {\bibfnamefont {S.}~\bibnamefont {Seiro}}, \bibinfo {author} {\bibfnamefont
  {C.}~\bibnamefont {Geibel}}, \bibinfo {author} {\bibfnamefont
  {S.}~\bibnamefont {Wirth}}, \bibinfo {author} {\bibfnamefont
  {F.}~\bibnamefont {Steglich}}, \ and\ \bibinfo {author} {\bibfnamefont
  {P.}~\bibnamefont {Wahl}},\ }\href {\doibase 10.1103/PhysRevB.93.045123}
  {\bibfield  {journal} {\bibinfo  {journal} {Phys. Rev. B}\ }\textbf {\bibinfo
  {volume} {93}},\ \bibinfo {pages} {045123} (\bibinfo {year}
  {2016})}\BibitemShut {NoStop}%
\bibitem [{\citenamefont {Kim}\ \emph {et~al.}(2017)\citenamefont {Kim},
  \citenamefont {Yoshida}, \citenamefont {Lee}, \citenamefont {Chang},
  \citenamefont {Jeng}, \citenamefont {Lin}, \citenamefont {Haga},
  \citenamefont {Fisk},\ and\ \citenamefont {Hasegawa}}]{Kim17}%
  \BibitemOpen
  \bibfield  {author} {\bibinfo {author} {\bibfnamefont {H.}~\bibnamefont
  {Kim}}, \bibinfo {author} {\bibfnamefont {Y.}~\bibnamefont {Yoshida}},
  \bibinfo {author} {\bibfnamefont {C.-C.}\ \bibnamefont {Lee}}, \bibinfo
  {author} {\bibfnamefont {T.-R.}\ \bibnamefont {Chang}}, \bibinfo {author}
  {\bibfnamefont {H.-T.}\ \bibnamefont {Jeng}}, \bibinfo {author}
  {\bibfnamefont {H.}~\bibnamefont {Lin}}, \bibinfo {author} {\bibfnamefont
  {Y.}~\bibnamefont {Haga}}, \bibinfo {author} {\bibfnamefont {Z.}~\bibnamefont
  {Fisk}}, \ and\ \bibinfo {author} {\bibfnamefont {Y.}~\bibnamefont
  {Hasegawa}},\ }\href@noop {} {\bibfield  {journal} {\bibinfo  {journal} {Sci.
  Adv.}\ }\textbf {\bibinfo {volume} {3}} (\bibinfo {year} {2017})}\BibitemShut
  {NoStop}%
\bibitem [{\citenamefont {Shishido}\ \emph {et~al.}(2010)\citenamefont
  {Shishido}, \citenamefont {Shibauchi}, \citenamefont {Yasu}, \citenamefont
  {Kato}, \citenamefont {Kontani}, \citenamefont {Terashima},\ and\
  \citenamefont {Matsuda}}]{Shishido10}%
  \BibitemOpen
  \bibfield  {author} {\bibinfo {author} {\bibfnamefont {H.}~\bibnamefont
  {Shishido}}, \bibinfo {author} {\bibfnamefont {T.}~\bibnamefont {Shibauchi}},
  \bibinfo {author} {\bibfnamefont {K.}~\bibnamefont {Yasu}}, \bibinfo {author}
  {\bibfnamefont {T.}~\bibnamefont {Kato}}, \bibinfo {author} {\bibfnamefont
  {H.}~\bibnamefont {Kontani}}, \bibinfo {author} {\bibfnamefont
  {T.}~\bibnamefont {Terashima}}, \ and\ \bibinfo {author} {\bibfnamefont
  {Y.}~\bibnamefont {Matsuda}},\ }\href {\doibase 10.1126/science.1183376}
  {\bibfield  {journal} {\bibinfo  {journal} {Science}\ }\textbf {\bibinfo
  {volume} {327}},\ \bibinfo {pages} {980} (\bibinfo {year}
  {2010})}\BibitemShut {NoStop}%
\bibitem [{\citenamefont {Shimozawa}\ \emph {et~al.}(2016)\citenamefont
  {Shimozawa}, \citenamefont {Goh}, \citenamefont {Shibauchi},\ and\
  \citenamefont {Matsuda}}]{Shimozawa16}%
  \BibitemOpen
  \bibfield  {author} {\bibinfo {author} {\bibfnamefont {M.}~\bibnamefont
  {Shimozawa}}, \bibinfo {author} {\bibfnamefont {S.~K.}\ \bibnamefont {Goh}},
  \bibinfo {author} {\bibfnamefont {T.}~\bibnamefont {Shibauchi}}, \ and\
  \bibinfo {author} {\bibfnamefont {Y.}~\bibnamefont {Matsuda}},\ }\href
  {http://stacks.iop.org/0034-4885/79/i=7/a=074503} {\bibfield  {journal}
  {\bibinfo  {journal} {Rep. Prog. Phys.}\ }\textbf {\bibinfo {volume} {79}},\
  \bibinfo {pages} {074503} (\bibinfo {year} {2016})}\BibitemShut {NoStop}%
\bibitem [{\citenamefont {Mizukami}\ \emph {et~al.}(2011)\citenamefont
  {Mizukami}, \citenamefont {Shishido}, \citenamefont {Shibauchi},
  \citenamefont {Shimozawa}, \citenamefont {Yasumoto}, \citenamefont
  {Watanabe}, \citenamefont {Yamashita}, \citenamefont {Ikeda}, \citenamefont
  {Terashima}, \citenamefont {Kontani},\ and\ \citenamefont
  {Matsuda}}]{Mizukami11}%
  \BibitemOpen
  \bibfield  {author} {\bibinfo {author} {\bibfnamefont {Y.}~\bibnamefont
  {Mizukami}}, \bibinfo {author} {\bibfnamefont {H.}~\bibnamefont {Shishido}},
  \bibinfo {author} {\bibfnamefont {T.}~\bibnamefont {Shibauchi}}, \bibinfo
  {author} {\bibfnamefont {M.}~\bibnamefont {Shimozawa}}, \bibinfo {author}
  {\bibfnamefont {S.}~\bibnamefont {Yasumoto}}, \bibinfo {author}
  {\bibfnamefont {D.}~\bibnamefont {Watanabe}}, \bibinfo {author}
  {\bibfnamefont {M.}~\bibnamefont {Yamashita}}, \bibinfo {author}
  {\bibfnamefont {H.}~\bibnamefont {Ikeda}}, \bibinfo {author} {\bibfnamefont
  {T.}~\bibnamefont {Terashima}}, \bibinfo {author} {\bibfnamefont
  {H.}~\bibnamefont {Kontani}}, \ and\ \bibinfo {author} {\bibfnamefont
  {Y.}~\bibnamefont {Matsuda}},\ }\href {\doibase 10.1038/nphys2112} {\bibfield
   {journal} {\bibinfo  {journal} {Nat. Phys.}\ }\textbf {\bibinfo {volume}
  {7}},\ \bibinfo {pages} {849} (\bibinfo {year} {2011})}\BibitemShut {NoStop}%
\bibitem [{\citenamefont {Goh}\ \emph {et~al.}(2012)\citenamefont {Goh},
  \citenamefont {Mizukami}, \citenamefont {Shishido}, \citenamefont {Watanabe},
  \citenamefont {Yasumoto}, \citenamefont {Shimozawa}, \citenamefont
  {Yamashita}, \citenamefont {Terashima}, \citenamefont {Yanase}, \citenamefont
  {Shibauchi}, \citenamefont {Buzdin},\ and\ \citenamefont {Matsuda}}]{Goh12}%
  \BibitemOpen
  \bibfield  {author} {\bibinfo {author} {\bibfnamefont {S.~K.}\ \bibnamefont
  {Goh}}, \bibinfo {author} {\bibfnamefont {Y.}~\bibnamefont {Mizukami}},
  \bibinfo {author} {\bibfnamefont {H.}~\bibnamefont {Shishido}}, \bibinfo
  {author} {\bibfnamefont {D.}~\bibnamefont {Watanabe}}, \bibinfo {author}
  {\bibfnamefont {S.}~\bibnamefont {Yasumoto}}, \bibinfo {author}
  {\bibfnamefont {M.}~\bibnamefont {Shimozawa}}, \bibinfo {author}
  {\bibfnamefont {M.}~\bibnamefont {Yamashita}}, \bibinfo {author}
  {\bibfnamefont {T.}~\bibnamefont {Terashima}}, \bibinfo {author}
  {\bibfnamefont {Y.}~\bibnamefont {Yanase}}, \bibinfo {author} {\bibfnamefont
  {T.}~\bibnamefont {Shibauchi}}, \bibinfo {author} {\bibfnamefont {A.~I.}\
  \bibnamefont {Buzdin}}, \ and\ \bibinfo {author} {\bibfnamefont
  {Y.}~\bibnamefont {Matsuda}},\ }\href {\doibase
  10.1103/PhysRevLett.109.157006} {\bibfield  {journal} {\bibinfo  {journal}
  {Phys. Rev. Lett.}\ }\textbf {\bibinfo {volume} {109}},\ \bibinfo {pages}
  {157006} (\bibinfo {year} {2012})}\BibitemShut {NoStop}%
\bibitem [{\citenamefont {Yamanaka}\ \emph {et~al.}(2015)\citenamefont
  {Yamanaka}, \citenamefont {Shimozawa}, \citenamefont {Endo}, \citenamefont
  {Mizukami}, \citenamefont {Shishido}, \citenamefont {Terashima},
  \citenamefont {Shibauchi}, \citenamefont {Matsuda},\ and\ \citenamefont
  {Ishida}}]{Yamanaka15}%
  \BibitemOpen
  \bibfield  {author} {\bibinfo {author} {\bibfnamefont {T.}~\bibnamefont
  {Yamanaka}}, \bibinfo {author} {\bibfnamefont {M.}~\bibnamefont {Shimozawa}},
  \bibinfo {author} {\bibfnamefont {R.}~\bibnamefont {Endo}}, \bibinfo {author}
  {\bibfnamefont {Y.}~\bibnamefont {Mizukami}}, \bibinfo {author}
  {\bibfnamefont {H.}~\bibnamefont {Shishido}}, \bibinfo {author}
  {\bibfnamefont {T.}~\bibnamefont {Terashima}}, \bibinfo {author}
  {\bibfnamefont {T.}~\bibnamefont {Shibauchi}}, \bibinfo {author}
  {\bibfnamefont {Y.}~\bibnamefont {Matsuda}}, \ and\ \bibinfo {author}
  {\bibfnamefont {K.}~\bibnamefont {Ishida}},\ }\href {\doibase
  10.1103/PhysRevB.92.241105} {\bibfield  {journal} {\bibinfo  {journal} {Phys.
  Rev. B}\ }\textbf {\bibinfo {volume} {92}},\ \bibinfo {pages} {241105}
  (\bibinfo {year} {2015})}\BibitemShut {NoStop}%
\bibitem [{\citenamefont {Shimozawa}\ \emph {et~al.}(2012)\citenamefont
  {Shimozawa}, \citenamefont {Watashige}, \citenamefont {Yasumoto},
  \citenamefont {Mizukami}, \citenamefont {Nakamura}, \citenamefont {Shishido},
  \citenamefont {Goh}, \citenamefont {Terashima}, \citenamefont {Shibauchi},\
  and\ \citenamefont {Matsuda}}]{Shimozawa12}%
  \BibitemOpen
  \bibfield  {author} {\bibinfo {author} {\bibfnamefont {M.}~\bibnamefont
  {Shimozawa}}, \bibinfo {author} {\bibfnamefont {T.}~\bibnamefont
  {Watashige}}, \bibinfo {author} {\bibfnamefont {S.}~\bibnamefont {Yasumoto}},
  \bibinfo {author} {\bibfnamefont {Y.}~\bibnamefont {Mizukami}}, \bibinfo
  {author} {\bibfnamefont {M.}~\bibnamefont {Nakamura}}, \bibinfo {author}
  {\bibfnamefont {H.}~\bibnamefont {Shishido}}, \bibinfo {author}
  {\bibfnamefont {S.~K.}\ \bibnamefont {Goh}}, \bibinfo {author} {\bibfnamefont
  {T.}~\bibnamefont {Terashima}}, \bibinfo {author} {\bibfnamefont
  {T.}~\bibnamefont {Shibauchi}}, \ and\ \bibinfo {author} {\bibfnamefont
  {Y.}~\bibnamefont {Matsuda}},\ }\href {\doibase 10.1103/PhysRevB.86.144526}
  {\bibfield  {journal} {\bibinfo  {journal} {Phys. Rev. B}\ }\textbf {\bibinfo
  {volume} {86}},\ \bibinfo {pages} {144526} (\bibinfo {year}
  {2012})}\BibitemShut {NoStop}%
\bibitem [{\citenamefont {Zhang}\ \emph {et~al.}(2009)\citenamefont {Zhang},
  \citenamefont {Cheng}, \citenamefont {Chen}, \citenamefont {Jia},
  \citenamefont {Ma}, \citenamefont {He}, \citenamefont {Wang}, \citenamefont
  {Zhang}, \citenamefont {Dai}, \citenamefont {Fang}, \citenamefont {Xie},\
  and\ \citenamefont {Xue}}]{Zhang09}%
  \BibitemOpen
  \bibfield  {author} {\bibinfo {author} {\bibfnamefont {T.}~\bibnamefont
  {Zhang}}, \bibinfo {author} {\bibfnamefont {P.}~\bibnamefont {Cheng}},
  \bibinfo {author} {\bibfnamefont {X.}~\bibnamefont {Chen}}, \bibinfo {author}
  {\bibfnamefont {J.-F.}\ \bibnamefont {Jia}}, \bibinfo {author} {\bibfnamefont
  {X.}~\bibnamefont {Ma}}, \bibinfo {author} {\bibfnamefont {K.}~\bibnamefont
  {He}}, \bibinfo {author} {\bibfnamefont {L.}~\bibnamefont {Wang}}, \bibinfo
  {author} {\bibfnamefont {H.}~\bibnamefont {Zhang}}, \bibinfo {author}
  {\bibfnamefont {X.}~\bibnamefont {Dai}}, \bibinfo {author} {\bibfnamefont
  {Z.}~\bibnamefont {Fang}}, \bibinfo {author} {\bibfnamefont {X.}~\bibnamefont
  {Xie}}, \ and\ \bibinfo {author} {\bibfnamefont {Q.-K.}\ \bibnamefont
  {Xue}},\ }\href {\doibase 10.1103/PhysRevLett.103.266803} {\bibfield
  {journal} {\bibinfo  {journal} {Phys. Rev. Lett.}\ }\textbf {\bibinfo
  {volume} {103}},\ \bibinfo {pages} {266803} (\bibinfo {year}
  {2009})}\BibitemShut {NoStop}%
\bibitem [{\citenamefont {Cheng}\ \emph {et~al.}(2010)\citenamefont {Cheng},
  \citenamefont {Song}, \citenamefont {Zhang}, \citenamefont {Zhang},
  \citenamefont {Wang}, \citenamefont {Jia}, \citenamefont {Wang},
  \citenamefont {Wang}, \citenamefont {Zhu}, \citenamefont {Chen},
  \citenamefont {Ma}, \citenamefont {He}, \citenamefont {Wang}, \citenamefont
  {Dai}, \citenamefont {Fang}, \citenamefont {Xie}, \citenamefont {Qi},
  \citenamefont {Liu}, \citenamefont {Zhang},\ and\ \citenamefont
  {Xue}}]{Cheng10}%
  \BibitemOpen
  \bibfield  {author} {\bibinfo {author} {\bibfnamefont {P.}~\bibnamefont
  {Cheng}}, \bibinfo {author} {\bibfnamefont {C.}~\bibnamefont {Song}},
  \bibinfo {author} {\bibfnamefont {T.}~\bibnamefont {Zhang}}, \bibinfo
  {author} {\bibfnamefont {Y.}~\bibnamefont {Zhang}}, \bibinfo {author}
  {\bibfnamefont {Y.}~\bibnamefont {Wang}}, \bibinfo {author} {\bibfnamefont
  {J.-F.}\ \bibnamefont {Jia}}, \bibinfo {author} {\bibfnamefont
  {J.}~\bibnamefont {Wang}}, \bibinfo {author} {\bibfnamefont {Y.}~\bibnamefont
  {Wang}}, \bibinfo {author} {\bibfnamefont {B.-F.}\ \bibnamefont {Zhu}},
  \bibinfo {author} {\bibfnamefont {X.}~\bibnamefont {Chen}}, \bibinfo {author}
  {\bibfnamefont {X.}~\bibnamefont {Ma}}, \bibinfo {author} {\bibfnamefont
  {K.}~\bibnamefont {He}}, \bibinfo {author} {\bibfnamefont {L.}~\bibnamefont
  {Wang}}, \bibinfo {author} {\bibfnamefont {X.}~\bibnamefont {Dai}}, \bibinfo
  {author} {\bibfnamefont {Z.}~\bibnamefont {Fang}}, \bibinfo {author}
  {\bibfnamefont {X.}~\bibnamefont {Xie}}, \bibinfo {author} {\bibfnamefont
  {X.-L.}\ \bibnamefont {Qi}}, \bibinfo {author} {\bibfnamefont {C.-X.}\
  \bibnamefont {Liu}}, \bibinfo {author} {\bibfnamefont {S.-C.}\ \bibnamefont
  {Zhang}}, \ and\ \bibinfo {author} {\bibfnamefont {Q.-K.}\ \bibnamefont
  {Xue}},\ }\href {\doibase 10.1103/PhysRevLett.105.076801} {\bibfield
  {journal} {\bibinfo  {journal} {Phys. Rev. Lett.}\ }\textbf {\bibinfo
  {volume} {105}},\ \bibinfo {pages} {076801} (\bibinfo {year}
  {2010})}\BibitemShut {NoStop}%
\bibitem [{\citenamefont {Wang}\ \emph {et~al.}(2017)\citenamefont {Wang},
  \citenamefont {Liu}, \citenamefont {Liu},\ and\ \citenamefont
  {Wang}}]{Wang17}%
  \BibitemOpen
  \bibfield  {author} {\bibinfo {author} {\bibfnamefont {Z.}~\bibnamefont
  {Wang}}, \bibinfo {author} {\bibfnamefont {C.}~\bibnamefont {Liu}}, \bibinfo
  {author} {\bibfnamefont {Y.}~\bibnamefont {Liu}}, \ and\ \bibinfo {author}
  {\bibfnamefont {J.}~\bibnamefont {Wang}},\ }\href@noop {} {\bibfield
  {journal} {\bibinfo  {journal} {J. Phys. Condens. Matter.}\ }\textbf
  {\bibinfo {volume} {29}},\ \bibinfo {pages} {153001} (\bibinfo {year}
  {2017})}\BibitemShut {NoStop}%
\bibitem [{\citenamefont {Georges}\ \emph {et~al.}(1996)\citenamefont
  {Georges}, \citenamefont {Kotliar}, \citenamefont {Krauth},\ and\
  \citenamefont {Rozenberg}}]{RevModPhys.68.13}%
  \BibitemOpen
  \bibfield  {author} {\bibinfo {author} {\bibfnamefont {A.}~\bibnamefont
  {Georges}}, \bibinfo {author} {\bibfnamefont {G.}~\bibnamefont {Kotliar}},
  \bibinfo {author} {\bibfnamefont {W.}~\bibnamefont {Krauth}}, \ and\ \bibinfo
  {author} {\bibfnamefont {M.~J.}\ \bibnamefont {Rozenberg}},\ }\href {\doibase
  10.1103/RevModPhys.68.13} {\bibfield  {journal} {\bibinfo  {journal} {Rev.
  Mod. Phys.}\ }\textbf {\bibinfo {volume} {68}},\ \bibinfo {pages} {13}
  (\bibinfo {year} {1996})}\BibitemShut {NoStop}%
\bibitem [{\citenamefont {Bulla}\ \emph {et~al.}(2008)\citenamefont {Bulla},
  \citenamefont {Costi},\ and\ \citenamefont {Pruschke}}]{RevModPhys.80.395}%
  \BibitemOpen
  \bibfield  {author} {\bibinfo {author} {\bibfnamefont {R.}~\bibnamefont
  {Bulla}}, \bibinfo {author} {\bibfnamefont {T.~A.}\ \bibnamefont {Costi}}, \
  and\ \bibinfo {author} {\bibfnamefont {T.}~\bibnamefont {Pruschke}},\ }\href
  {\doibase 10.1103/RevModPhys.80.395} {\bibfield  {journal} {\bibinfo
  {journal} {Rev. Mod. Phys.}\ }\textbf {\bibinfo {volume} {80}},\ \bibinfo
  {pages} {395} (\bibinfo {year} {2008})}\BibitemShut {NoStop}%
\bibitem [{\citenamefont {Nakajima}\ \emph {et~al.}(2007)\citenamefont
  {Nakajima}, \citenamefont {Shishido}, \citenamefont {Nakai}, \citenamefont
  {Shibauchi}, \citenamefont {Behnia}, \citenamefont {Izawa}, \citenamefont
  {Hedo}, \citenamefont {Uwatoko}, \citenamefont {Matsumoto}, \citenamefont
  {Settai}, \citenamefont {Onuki}, \citenamefont {Kontani},\ and\ \citenamefont
  {Matsuda}}]{Nakajima07}%
  \BibitemOpen
  \bibfield  {author} {\bibinfo {author} {\bibfnamefont {Y.}~\bibnamefont
  {Nakajima}}, \bibinfo {author} {\bibfnamefont {H.}~\bibnamefont {Shishido}},
  \bibinfo {author} {\bibfnamefont {H.}~\bibnamefont {Nakai}}, \bibinfo
  {author} {\bibfnamefont {T.}~\bibnamefont {Shibauchi}}, \bibinfo {author}
  {\bibfnamefont {K.}~\bibnamefont {Behnia}}, \bibinfo {author} {\bibfnamefont
  {K.}~\bibnamefont {Izawa}}, \bibinfo {author} {\bibfnamefont
  {M.}~\bibnamefont {Hedo}}, \bibinfo {author} {\bibfnamefont {Y.}~\bibnamefont
  {Uwatoko}}, \bibinfo {author} {\bibfnamefont {T.}~\bibnamefont {Matsumoto}},
  \bibinfo {author} {\bibfnamefont {R.}~\bibnamefont {Settai}}, \bibinfo
  {author} {\bibfnamefont {Y.}~\bibnamefont {Onuki}}, \bibinfo {author}
  {\bibfnamefont {H.}~\bibnamefont {Kontani}}, \ and\ \bibinfo {author}
  {\bibfnamefont {Y.}~\bibnamefont {Matsuda}},\ }\href {\doibase
  10.1143/JPSJ.76.024703} {\bibfield  {journal} {\bibinfo  {journal} {J. Phys.
  Soc. Jpn.}\ }\textbf {\bibinfo {volume} {76}},\ \bibinfo {pages} {024703}
  (\bibinfo {year} {2007})}\BibitemShut {NoStop}%
\bibitem [{\citenamefont {Jang}\ \emph {et~al.}(2014)\citenamefont {Jang},
  \citenamefont {White}, \citenamefont {Lum}, \citenamefont {Kim},
  \citenamefont {Tanatar}, \citenamefont {Straszheim}, \citenamefont
  {Prozorov}, \citenamefont {Keiber}, \citenamefont {Bridges}, \citenamefont
  {Shu}, \citenamefont {Baumbach}, \citenamefont {Janoschek},\ and\
  \citenamefont {Maple}}]{Jang14}%
  \BibitemOpen
  \bibfield  {author} {\bibinfo {author} {\bibfnamefont {S.}~\bibnamefont
  {Jang}}, \bibinfo {author} {\bibfnamefont {B.}~\bibnamefont {White}},
  \bibinfo {author} {\bibfnamefont {I.}~\bibnamefont {Lum}}, \bibinfo {author}
  {\bibfnamefont {H.}~\bibnamefont {Kim}}, \bibinfo {author} {\bibfnamefont
  {M.}~\bibnamefont {Tanatar}}, \bibinfo {author} {\bibfnamefont
  {W.}~\bibnamefont {Straszheim}}, \bibinfo {author} {\bibfnamefont
  {R.}~\bibnamefont {Prozorov}}, \bibinfo {author} {\bibfnamefont
  {T.}~\bibnamefont {Keiber}}, \bibinfo {author} {\bibfnamefont
  {F.}~\bibnamefont {Bridges}}, \bibinfo {author} {\bibfnamefont
  {L.}~\bibnamefont {Shu}}, \bibinfo {author} {\bibfnamefont {R.}~\bibnamefont
  {Baumbach}}, \bibinfo {author} {\bibfnamefont {M.}~\bibnamefont {Janoschek}},
  \ and\ \bibinfo {author} {\bibfnamefont {M.}~\bibnamefont {Maple}},\ }\href
  {\doibase 10.1080/14786435.2014.976287} {\bibfield  {journal} {\bibinfo
  {journal} {Philosophical Magazine}\ }\textbf {\bibinfo {volume} {94}},\
  \bibinfo {pages} {4219} (\bibinfo {year} {2014})}\BibitemShut {NoStop}%
\bibitem [{\citenamefont {Daniel}\ \emph {et~al.}(2005)\citenamefont {Daniel},
  \citenamefont {Bauer}, \citenamefont {Han}, \citenamefont {Booth},
  \citenamefont {Cornelius}, \citenamefont {Pagliuso},\ and\ \citenamefont
  {Sarrao}}]{Daniel05}%
  \BibitemOpen
  \bibfield  {author} {\bibinfo {author} {\bibfnamefont {M.}~\bibnamefont
  {Daniel}}, \bibinfo {author} {\bibfnamefont {E.~D.}\ \bibnamefont {Bauer}},
  \bibinfo {author} {\bibfnamefont {S.-W.}\ \bibnamefont {Han}}, \bibinfo
  {author} {\bibfnamefont {C.~H.}\ \bibnamefont {Booth}}, \bibinfo {author}
  {\bibfnamefont {A.~L.}\ \bibnamefont {Cornelius}}, \bibinfo {author}
  {\bibfnamefont {P.~G.}\ \bibnamefont {Pagliuso}}, \ and\ \bibinfo {author}
  {\bibfnamefont {J.~L.}\ \bibnamefont {Sarrao}},\ }\href {\doibase
  10.1103/PhysRevLett.95.016406} {\bibfield  {journal} {\bibinfo  {journal}
  {Phys. Rev. Lett.}\ }\textbf {\bibinfo {volume} {95}},\ \bibinfo {pages}
  {016406} (\bibinfo {year} {2005})}\BibitemShut {NoStop}%
\end{thebibliography}%

\end{document}